\documentclass[conference]{IEEEtran}
\IEEEoverridecommandlockouts
\usepackage{cite}
\usepackage{amsmath,amssymb,amsfonts}
\usepackage{algorithmic}
\usepackage{graphicx}
\usepackage{textcomp}
\usepackage{xcolor}
\usepackage{hyperref}
\usepackage{url}
\usepackage{booktabs}
\usepackage{tabularx}
\usepackage{enumitem} 
\usepackage{listings}

\ifCLASSOPTIONcompsoc
  \usepackage[caption=false,font=normalsize,labelfont=sf,textfont=sf]{subfig}
\else
  \usepackage[caption=false,font=footnotesize]{subfig}
\fi

\def\BibTeX{{\rm B\kern-.05em{\sc i\kern-.025em b}\kern-.08em
    T\kern-.1667em\lower.7ex\hbox{E}\kern-.125emX}}

\usepackage{fancyhdr}

\fancypagestyle{firstpage}{
  \fancyhf{} 
  
  \fancyhead[C]{\small 2025 34th International Conference on Parallel Architectures and Compilation Techniques (PACT)}
  
  \fancyfoot[C]{\begin{minipage}{\textwidth}\centering \scriptsize
    \copyright 2025 IEEE. Personal use of this material is permitted. Permission from IEEE must be obtained for all other uses, in any current or future media, including reprinting/republishing this material for advertising or promotional purposes, creating new collective works, for resale or redistribution to servers or lists, or reuse of any copyrighted component of this work in other works. \\
    \vspace{2mm}
    DOI: \href{https://doi.org/10.1109/PACT65351.2025.00027}{10.1109/PACT65351.2025.00027}
  \end{minipage}}
}

\begin{document}

\title{Agentic Auto-Scheduling: An Experimental Study of LLM-Guided Loop Optimization
}

\author{\IEEEauthorblockN{
Massinissa Merouani\IEEEauthorrefmark{1},
Islem Kara Bernou\IEEEauthorrefmark{1}, and
Riyadh Baghdadi\IEEEauthorrefmark{1}
}
\IEEEauthorblockA{\IEEEauthorrefmark{1}New York University Abu Dhabi, Abu Dhabi, UAE}
\IEEEauthorblockA{Corresponding authors: massinissa.m@nyu.edu; baghdadi@nyu.edu}
}

\newcommand{\medianAt}[1]{\FrameworkName@$#1$}
\newcommand{\BestOf}[2]{\FrameworkName$\_{#1}$@$#2$}

\newcommand{\smalltt}[1]{\texttt{\footnotesize #1}}

\definecolor{ColorYes}{HTML}{1b9e77} 
\definecolor{ColorNo}{HTML}{d95f02}  
\newcommand{\yes}{\textcolor{ColorYes}{Yes}}
\newcommand{\no}{\textcolor{ColorNo}{No}}

\newcommand{\FrameworkName}{\textsc{ComPilot}}
\newcommand{\FrameworkNameOrigin}{\textbf{Com}piler \textbf{Pilot}}

\maketitle
\thispagestyle{firstpage} 
\begin{abstract}
Automatic code optimization remains a difficult challenge, particularly for complex loop nests on modern hardware. This paper investigates a novel approach to code optimization where Large Language Models (LLMs) guide the process through a closed-loop interaction with a compiler. We present \FrameworkName{}, an experimental framework that leverages off-the-shelf LLMs, without any task-specific fine-tuning, as interactive optimization agents. \FrameworkName{} establishes a feedback loop where an LLM proposes transformations for a given loop nest to a compiler. The compiler attempts the transformations, reporting back legality status and measured speedup or slowdown. The LLM utilizes this concrete feedback to iteratively refine its optimization strategy. Our extensive evaluation across the PolyBench benchmark suite demonstrates the effectiveness of this zero-shot approach. \FrameworkName{} achieves geometric mean speedups of 2.66x (single run) and 3.54x (best-of-5 runs) over the original code. Furthermore, \FrameworkName{} demonstrates competitive performance against the state-of-the-art Pluto polyhedral optimizer, outperforming it in many cases. This experimental study demonstrates that general-purpose LLMs can effectively guide the code optimization process when grounded by compiler feedback, opening promising research directions for agentic AI in code optimization.
\end{abstract}


\section{Introduction}
Improving program performance remains a cornerstone of computer systems research and practice, directly impacting energy consumption, cloud computing costs, and the turnaround time for critical scientific and commercial applications. However, achieving optimal performance on modern hardware is notoriously difficult. The complex interplay of multi-level caches, sophisticated instruction pipelines, and diverse parallel execution resources creates a challenging landscape. Consequently, manual performance tuning requires prohibitive effort and expertise given the vastness of the search space of possible code transformations. Yet, automated compiler heuristics frequently struggle to deliver consistent results across today's diverse applications and hardware.

Loop nests are critical performance bottlenecks, particularly in scientific computing, image processing, and machine learning domains. Decades of research have yielded powerful optimization techniques, from compiler heuristics (e.g., in GCC~\cite{stallman2002gnu} and LLVM~\cite{lattner2004llvm}) and sophisticated polyhedral methods~\cite{Feautrier2011} (e.g., Pluto~\cite{bondhugula2008practical}) that analyze dependencies and orchestrate complex transformations (like tiling, fusion, parallelization), to empirical autotuning frameworks~\cite{ansel2014opentuner,8423171,Vuduc2011}. Despite these advances, consistently achieving optimal performance across diverse applications and hardware remains challenging, motivating the exploration of complementary optimization strategies.

Recent advancements in Large Language Models (LLMs) open up intriguing new possibilities. Beyond their well-known text generation capabilities, LLMs exhibit remarkable abilities in understanding complex instructions, reasoning about problems, and even analyzing source code structure. 
This potential has spurred recent investigations into using LLMs for code optimization tasks~\cite{cummins2024meta, cummins2023large, grubisic2024compiler, rosas2024should, shypula2023learning, PerfRL, taneja2025llm}. However, approaches generating optimized code directly~\cite{shypula2023learning, PerfRL, rosas2024should, taneja2025llm} often struggle to guarantee the semantic correctness of complex transformations without costly formal verification or brittle output checking.
Alternatively, other works focus on selecting compiler passes or flags~\cite{cummins2024meta, cummins2023large, grubisic2024compiler}. These methods are useful for orchestrating existing compiler heuristics, often at the Intermediate Representation (IR) level. However, they typically lack fine-grained control for composing sequences of high-level source transformations which are often essential for maximizing data locality and parallelism gains. Additionally, prior works often prioritized objectives other than execution speed, such as code size, and necessitated domain-specific LLM fine-tuning.
These challenges motivate exploring a more agentic paradigm, where the LLM acts as a proactive decision-maker interacting with its environment. This leads to our central research question: \textbf{Can off-the-shelf LLMs, grounded by empirical compiler feedback, effectively guide the complex process of loop optimization?}

\begin{figure*}[t]
    \centering
    \includegraphics[width=\textwidth]{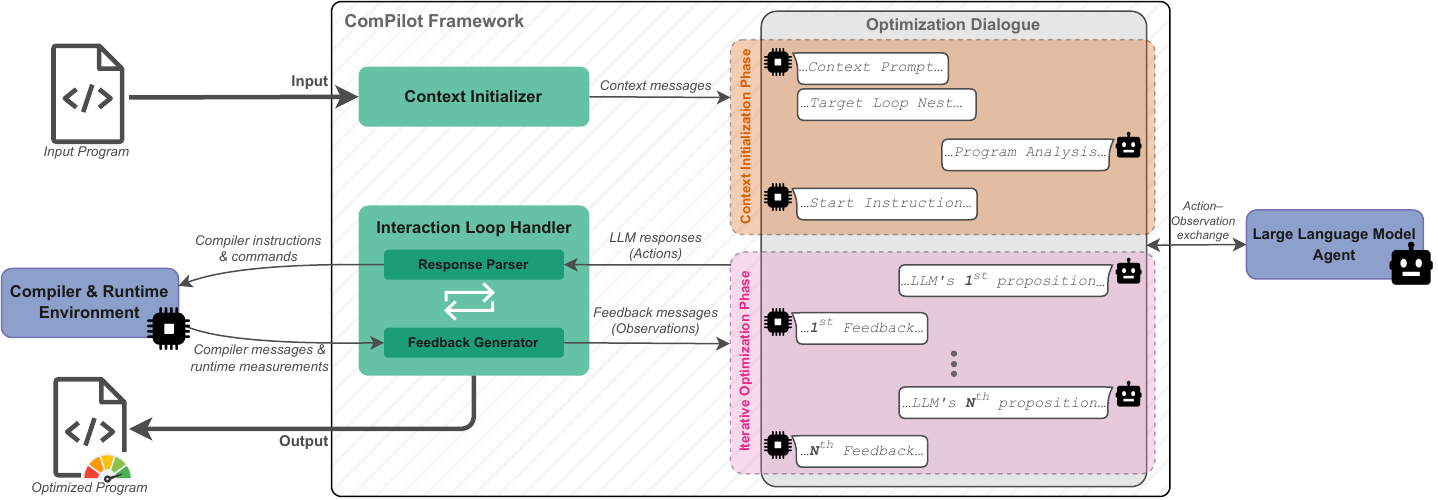} 
    \vspace{-0.5cm}
    \caption{Overview of the \FrameworkName{} framework, depicting the interaction between the LLM, the different modules, and the compiler.}
    \label{fig:system_overview}
    
\end{figure*}

To explore this question, we introduce \textbf{\FrameworkName{}} (\FrameworkNameOrigin{}), an experimental framework built around a closed-loop interaction between an LLM and a compiler infrastructure. In \FrameworkName{}, the LLM acts as an optimization agent, iteratively proposing sequences of loop transformations (schedules) for a given loop nest. These schedules are then passed to a compiler infrastructure (in our case, the Tiramisu compiler~\cite{tiramisu}) which attempts to apply the transformations, checks their legality based on dependence analysis~\cite{feautrier_array_1988,violateddep}, and generates code. \FrameworkName{} incorporates a feedback loop: the outcome of applying a schedule (success, failure type, and measured speedup/slowdown if successful) is reported back to the LLM. This closed-loop interaction allows the LLM to potentially learn from its successes and failures and refine its optimization strategy based on direct empirical evidence from the target machine. This approach leverages LLM capabilities to explore the transformation space while relying on the compiler's rigor for legality checking and code generation, all without requiring LLM fine-tuning or direct generation of transformed code.

This paper presents an experimental study evaluating the efficacy and characteristics of this agentic optimization approach. Our key contributions are:

\begin{itemize}
    \item The design and implementation of \FrameworkName{}, a framework enabling LLMs to interactively guide loop optimization using compiler-generated feedback.
    \item A demonstration of its effectiveness on PolyBench~\cite{louis-noel_polybench_2010}, showing that off-the-shelf LLMs can achieve significant speedups over strong baselines by discovering high-performance sequences of transformations.
    \item A detailed experimental study covering performance scaling, cost, variations across different LLMs, and ablation studies justifying the agentic design.
\end{itemize}

Our evaluation demonstrates the potential of LLM-guided optimization. Using a best-of-5 runs strategy, our approach achieves a geometric mean speedup of 3.54$\times$ over the original code and 2.94$\times$ over the state-of-the-art Pluto polyhedral compiler across the PolyBench suite~\cite{louis-noel_polybench_2010}. On certain benchmarks, \FrameworkName{} discovers schedules yielding speedups exceeding 100$\times$ compared to the original code.

\section{System Design and Methodology}
\label{sec:methodology}

We implement the idea of LLM-assisted loop optimization following the methodology described in Figure~\ref{fig:system_overview}.  We will first provide a high-level overview and then delve into the specifics of each component and the reasoning behind our design choices later in this section.

At the core of our system, \FrameworkName{} facilitates an interaction where the LLM acts as an agent within the compiler environment. For any given input program, this interaction is structured as an optimization dialogue—a dynamic conversation thread that serves two key functions in this agentic model: it is the interface for the agent's perception and action, and its history serves as the agent's episodic memory. This memory enables the agent to adapt its optimization strategy based on the concrete outcomes of past actions. This dialogue proceeds in iterations until specific stopping criteria are met. In each back-and-forth exchange, the LLM proposes loop transformations, and the compiler provides feedback on their validity, legality, and performance impact.

The process is composed of two main phases: the \textit{context initialization phase} and the \textit{iterative optimization phase}.

During the \textit{context initialization phase}, managed by the \textit{context initializer module}, the system briefs the LLM on its task and explains the overall flow of the optimization process through a context prompt (or a system instruction). The \textit{context initializer module} then extracts the relevant loop nest from the input program, presenting it to the LLM as a C/C++ style loop nest in a standardized format. The LLM is then prompted to analyze the loop nest. Once the analysis is complete, the LLM is directed to begin the optimization phase.

Once the context is initialized, the \textit{Iterative Optimization Phase} starts and the \textit{interaction loop handler} module takes over. This module processes the LLM's proposed optimizations (or schedules) by translating them into commands for the compiler or runtime environment. It then generates feedback to the LLM based on the execution outcomes, informing it about the schedule's validity, legality, and performance impact. This iterative exchange continues until a predefined stopping criterion is met. The system then outputs the optimized program variant with the best-achieved performance.


\subsection{Context Initialization Phase}

This phase sets the stage for the optimization dialogue through three key interactions between the \textit{Context Initializer} and the LLM.

\begin{figure}[h]
    \centering
    \includegraphics[width=\linewidth]{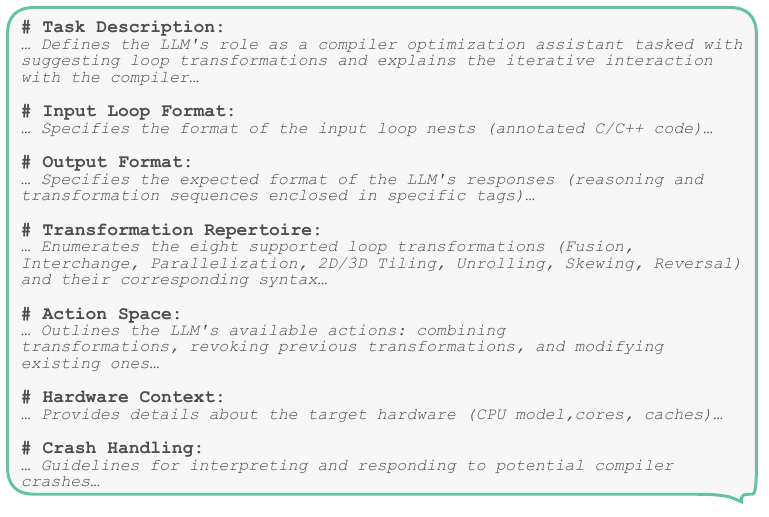} 
    \vspace{-1cm}
    \caption{Structure of the \textit{Context Prompt}}
    \label{fig:system_prompt_structure}
    
\end{figure}

First, the \textbf{Context Prompt} message is sent to the LLM at the beginning of each optimization session as system instructions. This message is constant across all input programs.  Its purpose is to clearly define the LLM's role as a compiler optimization assistant and to specify its expected behavior within the iterative optimization process. This prompt is structured as shown in Figure~\ref{fig:system_prompt_structure}, and the complete system prompt is provided in the appendix. This prompt outlines the process flow, input and output formats, transformations repertoire and action space, hardware target, and notes on how to handle errors and crashes. In this work, we experimented with a transformation space composed of nine primitives: Loop Fusion, Shifting, Interchange, Parallelization, 2D Tiling, 3D Tiling, Unrolling, Skewing, and Reversal. This choice of transformations provides a balance between expressiveness and manageability. We leverage Tiramisu's built-in solver for determining skewing and shifting factors when these transformations are involved, simplifying the LLM's task. The Fusion transformation might implicitly involve loop shifting operations to ensure legality.

\begin{figure}[h]
    \centering
    \includegraphics[width=\linewidth]{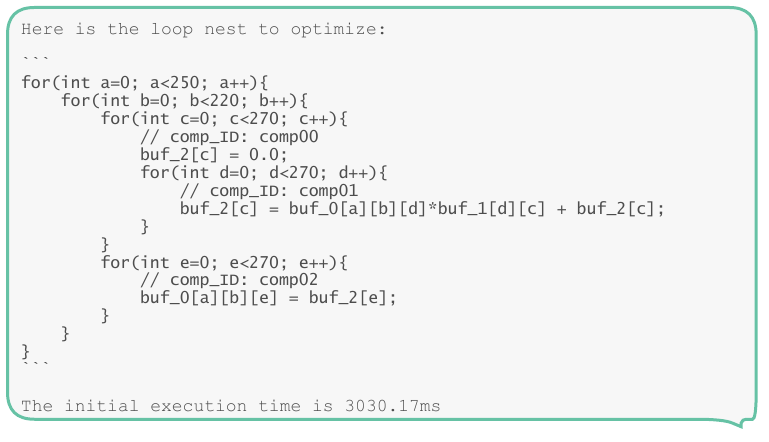} 
    \vspace{-0.8cm}
    \caption{Example of the message sent to the LLM showing the loop nest corresponding to the \smalltt{doitgen\_XLARGE} kernel.}
    \label{fig:loop_nest_example}
    
\end{figure}

Next, the \textit{Context Initializer Module} presents the specific \textbf{Target Loop Nest} to the LLM in a standardized manner (example in Figure~\ref{fig:loop_nest_example}). The \textit{Context Initializer Module} first extracts the relevant loop nest and annotates the C/C++ code with special comments (e.g., \smalltt{// comp\_ID: comp05}) to assign unique identifiers to each computation block. These \smalltt{comp\_ID}s allow the LLM to precisely specify the target of each transformation. It then anonymizes the loop iterator and buffer names by systematically replacing them with neutral identifiers (e.g., \smalltt{a}, \smalltt{b}, \smalltt{c}, \smalltt{buf0}, \smalltt{buf1}). 
This anonymization is performed for the sake of robustness, it prevents the LLM from being unduly influenced by potentially misleading or semantically irrelevant naming convention.
Along with the code, the loop nest's initial execution time is provided to the LLM, giving it a baseline performance measure for the target hardware.

\begin{figure}[h]
    \centering
    \includegraphics[width=\linewidth]{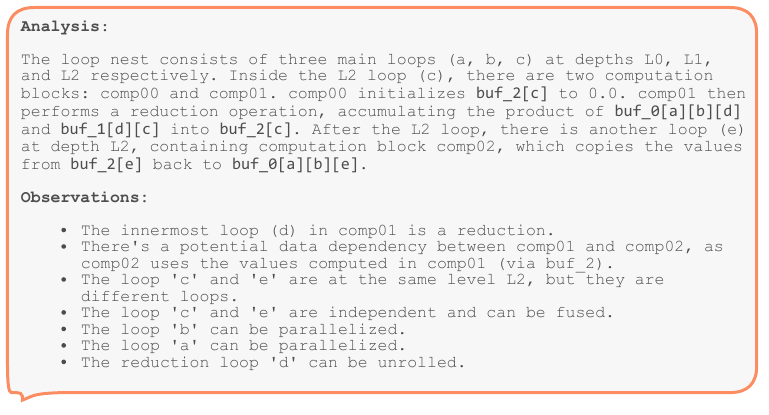}
    \vspace{-0.9cm}
    \caption{Example program analysis generated by the LLM.}
    \label{fig:llm_program_analysis}
    
\end{figure}

Finally, the \textit{Context Initializer Module} prompts the LLM to first \textbf{analyze the target loop nest} before proposing transformations. An example of such an analysis is shown in Figure~\ref{fig:llm_program_analysis}. This analysis serves as a form of chain-of-thought reasoning~\cite{NEURIPS2022_9d560961}, guiding all the LLM's subsequent optimization decisions. The LLM is expected to: deconstruct the loop nest's structure, infer the purpose of the loop nest and the role of individual computations, and propose potential optimization strategies based on its understanding. The importance of this program analysis phase is demonstrated experimentally in the appendix (RQ10).

With the context set and initial analysis complete, the \textit{Context Initializer Module} instructs the LLM to begin suggesting transformations, transitioning to the \textit{Iterative Optimization Phase.}

\subsection{Iterative Optimization Phase}

This phase constitutes the core of \FrameworkName{}'s iterative optimization process. During each iteration, the LLM reasons based on its current strategy and interaction history, and takes an action by proposing transformation candidates.  The \textit{Interaction Loop Handler} executes this action within the compiler environment, generating feedback that serves as an observation of the outcome. The LLM then uses this observation to update its strategy and plan its next action.

\begin{figure}[h]
    \centering
    \includegraphics[width=\linewidth]{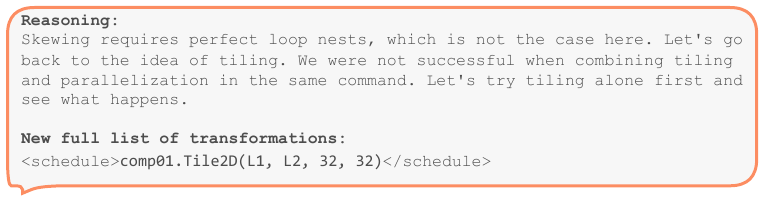}
    \vspace{-0.8cm}
    \caption{An example of the LLM's schedule proposition message.}
    \label{fig:llm_suggestion_parsing}
    
\end{figure}

Within the \textit{Interaction Loop Handler}, the \textit{Response Parser} submodule is responsible for processing the LLM's responses. As per the context prompt, the LLM is expected to provide a structured response (example in Figure~\ref{fig:llm_suggestion_parsing}). It must contain a \textbf{reasoning section},  explaining its rationale for the proposed transformation sequence based on its strategy and previous feedback. This section serves as an explicit chain-of-thought mechanism, the importance of which we verify experimentally in the appendix (RQ10). Following the reasoning, the LLM must provide the actual transformation sequence within \smalltt{<schedule>...</schedule>} tags using the syntax defined in the context prompt. Enforcing this structured output facilitates the systematic extraction of schedules by the \textit{Response Parser}.

 After extracting the LLM's proposed transformations, \FrameworkName{} performs a two-stage correctness check. First, a lightweight, compiler-independent validity check filters out syntactically malformed or semantically nonsensical propositions. This pre-filtering step verifies adherence to the transformation syntax, the use of valid identifiers, and fundamental preconditions (e.g., perfect nesting for loop interchange), thus preventing unnecessary and costly compiler interactions.

Schedules that pass this initial check proceed to a formal legality check performed by the backend compiler. In our implementation, Tiramisu employs rigorous polyhedral dependence analysis~\cite{feautrier_array_1988,violateddep} to guarantee that a transformation preserves the original program's semantics. This delegation is a core design principle: \FrameworkName{} leverages the LLM for high-level strategic exploration while entrusting the compiler with formal correctness, ensuring code reliability without brittle runtime output comparisons. For legal schedules, the \textit{Response Parser} emits the corresponding Tiramisu API calls to implement the transformations. When parameter calculations are required (e.g., skewing factors or fusion-induced shifting), \FrameworkName{} invokes Tiramisu’s internal solvers. The transformed code is then compiled and executed on the target machine to measure runtime and compute speedup or slowdown.

\begin{figure}[h]
    \centering
    \includegraphics[width=\linewidth]{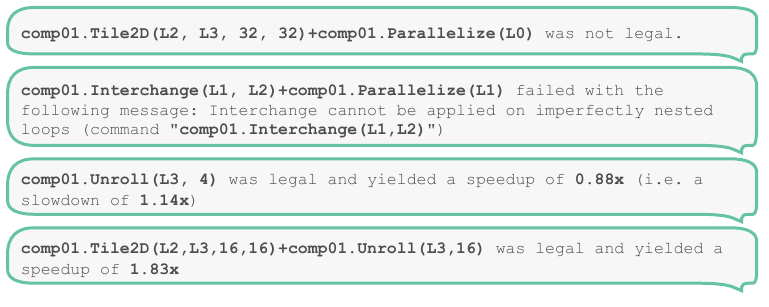}
    \vspace{-0.5cm}
    \caption{Examples of different feedback messages.}
    \label{fig:feedback_examples}
    
\end{figure}

The LLM is informed of the outcome of the parsing, compilation, and execution steps by the \textit{Feedback Generator} submodule. This submodule constructs a feedback message depending on the stage at which processing the LLM's suggestion succeeded or failed (examples in Figure~\ref{fig:feedback_examples}). We distinguish five categories of feedback:

\begin{itemize}[leftmargin=1.0em, labelsep=0.5em]
    \item \textbf{Invalid Schedule:} If the \textit{Response Parser} detects an invalid transformation, the feedback explains the specific reason for the invalidity.
    \item \textbf{Illegal Schedule:} If the compiler's legality checker determines that the schedule is illegal (violates data dependencies), the feedback message indicates so.
    \item \textbf{Solver Failure:} If Tiramisu's solvers cannot find valid parameters for skewing or shifting, the feedback reports so.
    \item \textbf{Compiler Crash:} If the compiler crashes (which generally occurs due to invalid transformations that are not detected by our set of rules), the feedback reports the crash, along with any available error messages.
    \item \textbf{Successful Execution:} If the schedule is valid, legal, and the transformed program executes successfully, the feedback provides the achieved speedup (or slowdown), calculated as the ratio of the original program's execution time to the transformed program's execution time.
\end{itemize}

This feedback, containing the outcome (success/failure type, performance metrics), is appended to the optimization history to update the LLM's working memory. For the next iteration, the LLM processes this entire updated context. This makes the LLM utilize its inherent in-context learning (ICL)~\cite{NEURIPS2020_1457c0d6} capabilities to interpret this feedback and the preceding interaction history, adapting its strategy and reasoning process to generate the subsequent schedule proposition. This mechanism allows \FrameworkName{} to leverage the LLM's adaptability dynamically within the dialogue, without requiring explicit fine-tuning or gradient updates, relying solely on the information provided in the prompt context.

The iterative process continues until a stopping condition is met. The LLM is instructed to issue the \smalltt{no\_further\_transformations} command when it believes no further promising transformations are available. However, our observations reveal that the LLM often exhibits a tendency to stop prematurely, either after a significant speedup jump (due to conservatism, wanting to avoid potentially detrimental transformations) or after repeated unsuccessful attempts (getting stuck in local optima). To mitigate premature stopping, the Interaction Loop Handler can prompt the LLM to continue exploring,  pushing it to consider additional transformations. The impact of continued exploration (RQ11) is analyzed in the appendix. Furthermore, to address the local optima problem, we employ a multi-run strategy (repeated trials), restarting the optimization dialogue from scratch multiple times (see RQ1 and RQ9). 

The optimization process terminates when either the LLM issues the stopping command and the framework chooses not to prompt further exploration, or a predefined iteration limit is reached.

\section{Results and Analysis}
\label{sec:results_analysis}

In this section, we evaluate our proposed LLM-assisted optimization process. Given the experimental nature of this work, we structure our analysis around a series of research questions (RQs). We first present the \textbf{main performance results} achieved by \FrameworkName{} under different scenarios and compare them against relevant baselines. We then detail \textbf{ablation studies} justifying key design choices. Further analyses characterizing performance scaling and the impact of specific interaction strategies are summarized in a concluding subsection, with full details provided in the appendix.


\vspace{0.5cm}
\noindent \textbf{\emph{Experimental Setup:}}~\begin{itemize}[leftmargin=1.0em, labelsep=0.25em]
    \item \textbf{Hardware:} All execution time measurements were performed on a dual-socket machine equipped with two 12-core Intel(R) Xeon(R) CPU E5-2695 v2 processors (@2.40GHz), totaling 48 threads, and 128GB of RAM.
    \item \textbf{Compiler:} We utilized the Tiramisu compiler~\cite{tiramisu} (commit \href{https://github.com/Tiramisu-Compiler/tiramisu/tree/041afadf050899e33695600db6241024b2f46088}{\smalltt{041afad}}) to perform legality checks, invoke its internal solvers (e.g., for skewing factors), apply the LLM-suggested transformation sequences, and generate executable code.
    \item \textbf{LLMs:} Unless otherwise specified, results are generated using \smalltt{gemini-2.0-flash}, chosen for its favorable balance of performance and inference cost at the time of our experiments. Comparisons with other prominent LLMs are presented in RQ4. Proprietary LLMs were accessed via cloud APIs while open-source ones were generally run on local hardware.
    \item \textbf{Benchmarks:} We evaluated \FrameworkName{} on the PolyBench/C benchmark suite \footnote{\url{https://polybench.sf.net}}~\cite{louis-noel_polybench_2010}(version 4.2.1), a standard benchmark set for polyhedral compilation research. PolyBench comprises 30 benchmarks from diverse domains (linear algebra, stencils, etc.). For each benchmark, we used all five standard dataset sizes and default data types, resulting in 150 distinct benchmark instances.
\end{itemize}

\vspace{0.1cm}
\noindent \textbf{\emph{Addressing LLM Stochasticity:}} 
LLM outputs are stochastic, so repeated \FrameworkName{} runs on the same program can yield different dialogues and final schedules. To obtain robust, representative estimates, we execute many independent runs per program across all 150 instances, forming a pool of results. For a single instance, we report the \textit{median speedup} over its pool (robust to outliers and reflective of central tendency). To aggregate across benchmarks, we take the \textit{geometric mean} of these medians—standard for ratios like speedups, as it weights relative improvements equally and is less skewed by extreme values than the arithmetic mean. To assess stability under run-to-run variability, we use \textit{bootstrapping} to compute 95\% confidence intervals by resampling each instance’s speedup pool; we report these CIs alongside all geometric-mean results. Further details on the bootstrapping procedure are provided in the appendix.

\vspace{0.1cm}
\noindent {\textbf{\emph{Metrics and Notation:}}} We evaluate \FrameworkName{}'s performance under two primary scenarios:
\begin{enumerate}[leftmargin=1.0em, labelsep=0.2em]
    \item \textbf{Single-Run:} This scenario answers: ``If a user runs \FrameworkName{} \textit{once} on a program, what speedup can they typically expect?"
    \item \textbf{Multi-Run (Best-of-$K$):} This scenario answers: ``If a user runs \FrameworkName{} $K$ times and selects the best result, what is the typical best speedup they can achieve?"
\end{enumerate}

Performance naturally depends on the \textbf{number of iterations ($T$)} allowed per run (i.e., the number of distinct schedules the LLM explores) and, for the multi-run scenario, the \textbf{number of runs ($K$)} performed. Longer explorations (larger $T$) generally allow the LLM to refine schedules and potentially discover better optima, while more runs (larger $K$) increase the diversity of explored schedules, improving the chance of finding a high-performing one. The trade-offs associated with $K$ and $T$ are analyzed in detail in RQ9.

\noindent We use the following notation:

\begin{itemize}[leftmargin=1.0em, labelsep=0.25em]
    \item \textbf{\medianAt{T}}: Represents the typical single-run speedup after $T$ iterations. For a single benchmark, this is the median speedup across the pool of runs, each stopped after $T$ iterations. When aggregated across benchmarks, it is the geometric mean of these medians.
    \item \textbf{\BestOf{K}{T}}: Represents the typical best-of-$K$ speedup after $T$ iterations per run. For a single benchmark, this is calculated by repeatedly sampling $K$ runs from the pool, finding the maximum speedup within each sample of $K$, and taking the median of these maximums. This simulates the typical best result a user would get by running \FrameworkName{} $K$ times. Aggregation across benchmarks uses the geometric mean of these values.
\end{itemize}

Note that by definition, \BestOf{1}{T} is equivalent to \medianAt{T}.

\subsection{Main Results}
\label{subsec:main_results}

\begin{figure*}[h]
    \centering
    \includegraphics[width=\textwidth]{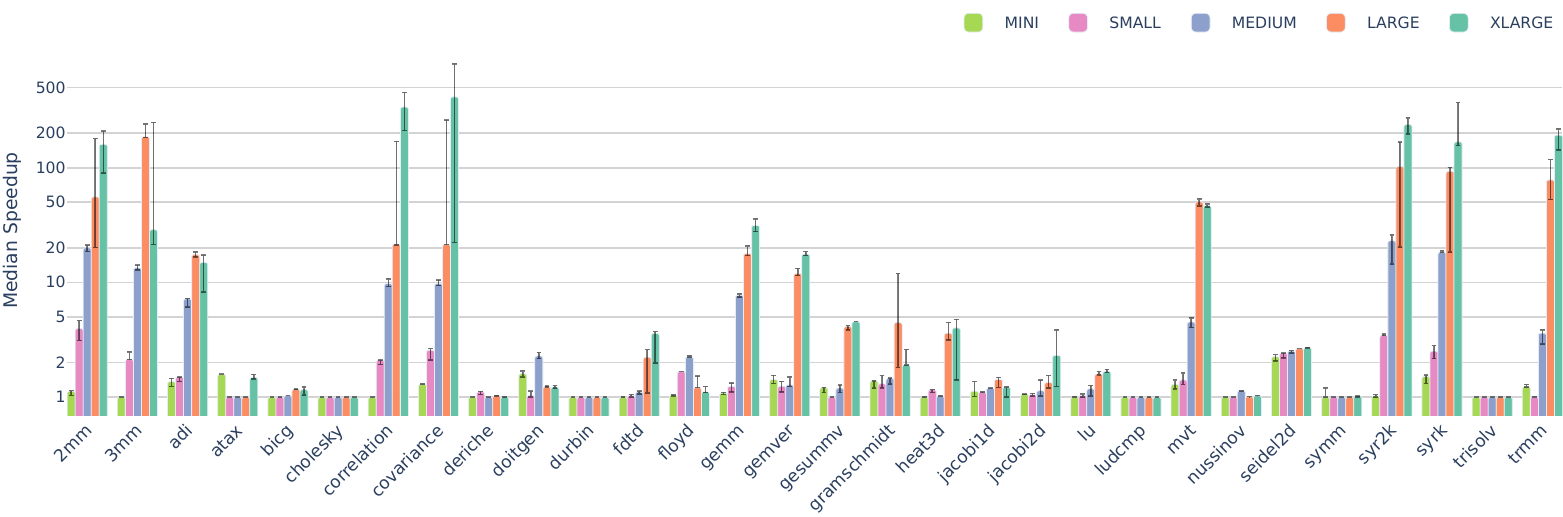}
    \vspace{-0.8cm}
    \caption{\medianAt{30} speedups per benchmark instance. Each bar represents the median speedup over the pool of 40 runs with error bars indicating the 95\% bootstrapped confidence interval for that median.}
    \vspace{-0.4cm}
    \label{fig:collm30_speedups} 
    
\end{figure*}

\subsubsection*{\textbf{RQ1: What are the typical speedups that \FrameworkName{} achieves?}}
\label{subsubsec:rq1}

To ensure stable results, our analysis is based on a pool of 40 independent runs for each of the 150 PolyBench instances. We focus here on performance after 30 iterations (@30), as RQ9 will show this provides a reasonable balance between optimization time and achieved speedup.

\paragraph*{\textbf{Single-Run Evaluation}}
This evaluates the typical speedup a user would achieve from a single execution of \FrameworkName{}. Figure~\ref{fig:collm30_speedups} presents the \medianAt{30} speedups for all 150 PolyBench instances. Each bar represents the median speedup achieved over the pool of runs with error bars indicating the 95\% bootstrapped confidence interval for that median.

Overall, \medianAt{30} achieves an aggregate geometric mean speedup of 2.66$\times$ across all 150 benchmark instances compared to the original unoptimized code. The 95\% confidence interval is [2.60, 2.77] for this geometric mean, indicating good stability despite the LLM's stochastic nature. The distribution of speedups shows that significant gains are common: \medianAt{30} achieves a median speedup of at least 1.24$\times$ in 50\% of the instances, at least 3.6$\times$ in 25\%, and exceeds 23.65$\times$ in the top 10\%.

Performance gains exhibit significant variability across benchmarks and input sizes, as detailed in Figure~\ref{fig:collm30_speedups}. Notably, larger input sizes often enable higher speedups. For instance, \smalltt{correlation\_XLARGE} attained a median speedup of 339$\times$, primarily by aggressively parallelizing multiple outer loops (\smalltt{comp00.Parallelize(L0)+...}) combined with tiling and unrolling (\smalltt{...+comp07.Tile2D(L1,L2,32,32)+...}), effectively leveraging the 48 available threads. Similarly, \smalltt{trmm\_XLARGE} reached 183$\times$ by using loop interchange to enable outer-loop parallelization (\smalltt{comp01.Interchange(L0,L1) + comp01.Parallelize(L0)+...}) along with tiling. This suggests \FrameworkName{} successfully prioritizes thread-level parallelism for large problems on the multi-core test system.

Conversely, for smaller inputs, \FrameworkName{} often favors locality or transformations enabling modest parallelism. \smalltt{trmm\_MEDIUM}, for example, achieved 3.6$\times$ via tiling and inner-loop parallelization (\smalltt{comp00.Tile2D(L0,L1,16,16) + comp00.Parallelize(L1)}), while \smalltt{seidel2d\_SMALL} saw a 2.41$\times$ speedup from skewing only (\smalltt{comp00.Skew(L1,L2)}). 
This indicates an adaptive optimization strategy sensitive to input scale. The complexity of the final schedules found by \FrameworkName{} also varied considerably, from single transformations to lengthy sequences, reflecting the exploratory nature of the dialogue.

However, several benchmarks (e.g., \smalltt{cholesky}, \smalltt{durbin}, \smalltt{ludcmp}) showed negligible improvement (median speedup $\approx 1\times$) across all input sizes. We hypothesize this stems from complex loop-carried dependencies that prove challenging for the LLM to optimize effectively using the current set of transformation primitives. Exploring transformations beyond the current set, such as loop distribution or computation reordering, might be necessary for these cases.

\begin{figure*}[t]
    \centering
    \includegraphics[width=\linewidth]{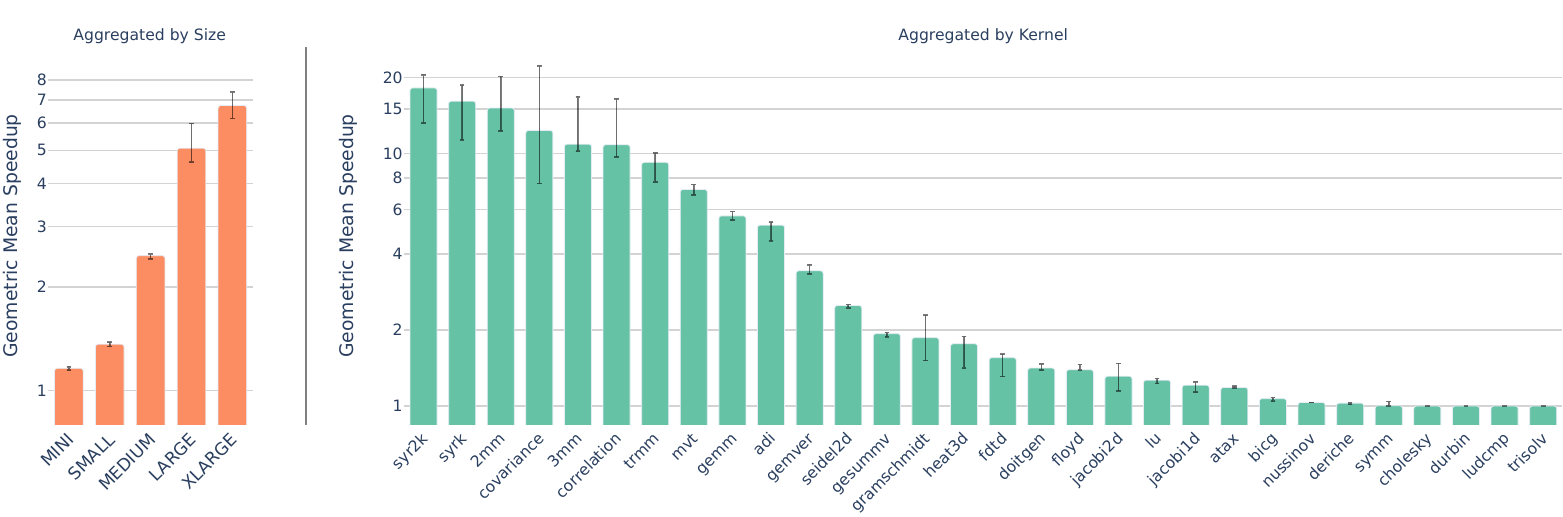}
    \vspace{-0.4cm}
    \caption{Geometric mean aggregate of \medianAt{30} speedups by input size (left) and benchmark kernel (right). Error bars indicate 95\% CIs.}
    \label{fig:collm30_speedup_aggregates}
    
\end{figure*}

Figure~\ref{fig:collm30_speedup_aggregates} shows speedups aggregated by input size and benchmark kernel, respectively, further illustrating these trends. Some benchmarks exhibit large confidence intervals in Figure~\ref{fig:collm30_speedups}, suggesting the LLM converges to different local optima across runs, resulting in a multi-modal distribution of speedups. We provide visualizations of these per-run distributions in the appendix. 

\paragraph*{\textbf{Multi-Run Evaluation (Best-of-K)}}
This evaluates the typical best speedup achievable if a user runs \FrameworkName{} multiple times. We focus here on $K=5$ runs (\BestOf{5}{30}), a choice further discussed in RQ9 as offering diminishing returns beyond this point.

By selecting the best schedule from 5 independent runs, \BestOf{5}{30} achieves an aggregate geometric mean speedup of 3.54$\times$. The 95\% confidence interval is [3.45, 3.58]. As expected, this is significantly higher than the single-run performance, demonstrating the benefit of exploring diverse optimization paths offered by the LLM's stochasticity. With 5 runs, the median speedup (\BestOf{5}{30}) is at least 1.59$\times$ in 50\% of instances, at least 4.8$\times$ in 25\%, and exceeds 53.65$\times$ in the top 10\%. This highlights that multiple runs substantially increase the likelihood of discovering highly effective optimization schedules. A bar chart of per-benchmark speedups of \BestOf{5}{30} is provided in the appendix.

\subsubsection*{\textbf{RQ2: What are the typical runtime and token consumption of \FrameworkName{}?}}
\label{subsubsec:RQ2_}
\begin{figure}[t]
    \centering
    \includegraphics[width=0.8\linewidth]{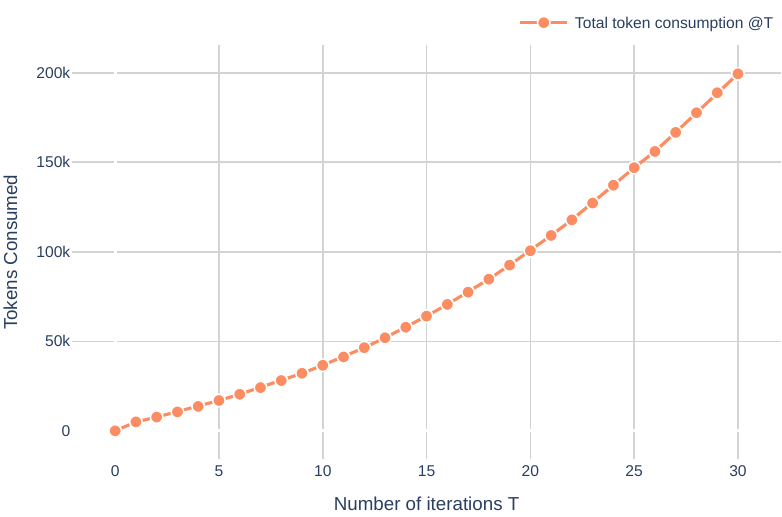}
    \caption{Average token consumption of \medianAt{T} as a function of the number of iterations (T)}
    \label{fig:token_evolution}
    
\end{figure}

We evaluate \FrameworkName{}'s operational cost in terms of runtime and LLM token consumption. To quantify token usage, we cumulate the number of tokens consumed in all iterations up to a given iteration, reflecting the total number of tokens consumed to that point. 
Figure~\ref{fig:token_evolution} illustrates the average cumulative token usage as a function of the number of iterations $T$. The values shown represent the total token usage (sum of input and output tokens).

The plot shows that token usage grows non-linearly. Two key factors contribute to this acceleration. Firstly, since LLM interactions typically require sending the entire preceding dialogue history, each subsequent turn consumes more tokens simply due to the growing context size. Secondly, our definition counts an ``iteration" only upon the exploration of a novel schedule. Especially in later stages, the LLM may propose repeated suggestions or attempt premature termination, requiring several unproductive exchanges (each consuming tokens) before a new schedule is explored. This significantly increases the total token cost associated with reaching higher iteration counts.

In terms of wall-clock time, running \FrameworkName{} for 30 iterations averaged approximately 8.9 minutes per benchmark instance on our setup. This runtime varies significantly by input size, ranging from an average of 16 minutes for \smalltt{XLARGE} instances down to roughly 5-6 minutes for \smalltt{MEDIUM}, \smalltt{SMALL}, and \smalltt{MINI} sizes. Analyzing the runtime breakdown reveals that direct communication with the LLM is not the bottleneck. Using \smalltt{gemini-2.0-flash}, the total communication time is 1-3 minutes per benchmark, although this specific timing is subject to variability depending on the LLM used and API provider performance. The majority of the processing time, around 78.5\% in our setup, is consumed by the backend compiler infrastructure. This time is spent checking schedule legality, compiling the resulting code, and (for the most part) executing these transformed versions to gather performance data.

\subsubsection*{\textbf{RQ3: How often are the LLM's schedule propositions valid and legal?}}
\label{subsubsec:RQ3} 

As described in Section~\ref{sec:methodology}, not every transformation sequence proposed by the LLM leads to runnable code. Some of the schedules can be invalid (syntactically/semantically flawed) or illegal (violating dependencies). Overall, considering dialogues up to 30 iterations ($T=30$) averaged across all runs and benchmarks, we find that 36.1\% of the proposed schedules are successfully compiled and run (\textit{runnable}), while 31.4\% are \textit{invalid} and 32.5\% are \textit{illegal}. Thus, roughly two-thirds of the LLM's propositions are unproductive attempts, highlighting a key challenge in using LLMs for this task. 
Interestingly, these proportions are not static throughout the dialogue. Illegal propositions are significantly more prevalent in the initial iterations (nearly 60\% at $T=1$), but this rate decreases as the dialogue progresses, potentially indicating the LLM learns from negative feedback. A visualization and discussion regarding this behavior are provided in the appendix.

\begin{figure}[t]
    \centering
    \includegraphics[width=\linewidth]{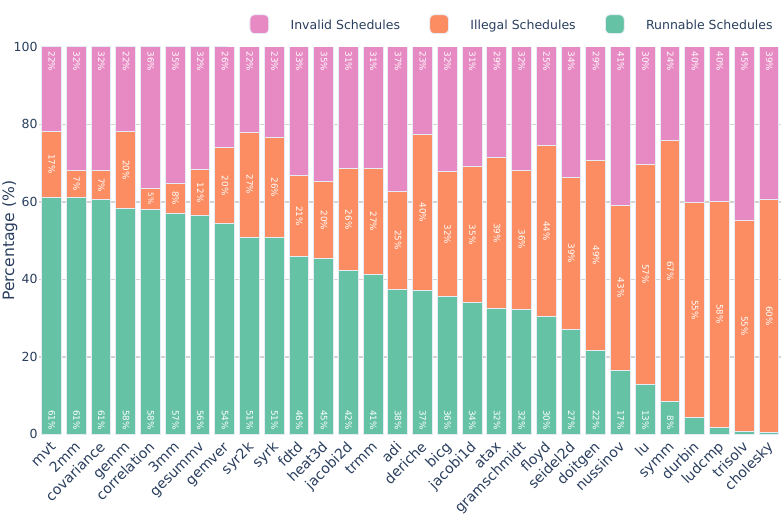} 
    \vspace{-0.8cm}
    \caption{Breakdown of proposed schedule types (runnable, invalid, illegal) per kernel at iteration T=30.}
    \label{fig:exploration_ratios_per_bench}
    
\end{figure}

These average ratios vary significantly across benchmarks (Figure~\ref{fig:exploration_ratios_per_bench}). For benchmarks like \smalltt{mvt}, \smalltt{2mm}, and \smalltt{covariance}, over 60\% of proposed schedules are runnable, indicating relatively easier optimization spaces for the LLM to navigate. Conversely, for kernels with complex dependencies like \smalltt{cholesky}, \smalltt{durbin}, and \smalltt{ludcmp}, fewer than 5\% of propositions result in runnable code, with illegality being the dominant failure mode. This strongly correlates with the performance results in RQ1, where these specific benchmarks saw little to no speedup. The ratio of invalid schedules shows less pronounced variation across benchmarks, suggesting these errors are less tied to inherent program complexity and more related to the LLM's general ability to adhere to the transformation syntax and rules.

It is also important to note that these ratios are LLM-dependent. The results presented above are computed for our primary LLM (\smalltt{gemini-2.0-flash}). 
our experiments with other models yielded different distributions (e.g., \smalltt{codestral-2501-22B}: 64.5\% invalid, 20.5\% illegal, 15\% runnable). This is further discussed in RQ4.

\subsubsection*{\textbf{RQ4: How Does \FrameworkName{} Perform with Different LLMs?}}
\label{subsubsec:RQ4}

The choice of the underlying LLM can significantly influence \FrameworkName{}'s effectiveness. To investigate this, we evaluated \FrameworkName{}'s performance using a selection of eight LLMs prominent at the time of writing. This selection aimed for diversity, including proprietary and open-source models, models with reasoning capability versus standard non-reasoning ones, and models focused on coding tasks versus general-purpose tasks. 

Table~\ref{tab:llm_speedups_single} presents the single-run (\medianAt{T}) geometric mean speedups for these LLMs at various iteration counts ($T$). Best results per column within 5\% tolerance are bolded. A similar table for the multi-run scenario (\BestOf{5}{T}) is provided in the appendix.

\begin{table}[t]
\caption{\medianAt{T} geomean across all benchmarks for different LLMs at various iteration (\textit{T}). N/E: not evaluated.}
\label{tab:llm_speedups_single}
\centering
\footnotesize
\setlength{\tabcolsep}{4pt} 
\begin{tabular}{lcccccc}
    \toprule
    LLM & \emph{T}=5 & \emph{T}=10 & \emph{T}=15 & \emph{T}=20 & \emph{T}=25 & \emph{T}=30 \\
    \midrule
    \texttt{gemini-2.0-flash} & 1.83 & 2.06 & \textbf{2.32} & \textbf{2.49} & \textbf{2.58} & \textbf{2.66} \\
    \texttt{gemma3 (27B)}    & 1.67 & 1.84 & 1.92 & 1.96 & 2.02 & 2.03 \\
    \texttt{gpt-4o}            & \textbf{1.98} & \textbf{2.26} & \textbf{2.39} & \textbf{2.51} & \textbf{2.57} & \textbf{2.63} \\
    \texttt{llama3.3 (70B)}  & 1.86 & 2.11 & 2.22 & 2.33 & 2.40 & 2.47 \\
    \texttt{gpt-o3-mini}      & \textbf{1.99} & \textbf{2.25} & \textbf{2.42} & \textbf{2.51} & \textbf{2.63} & \smalltt{N/E} \\
    \texttt{qwq (32B)}           & \textbf{2.02} & \textbf{2.21} & \textbf{2.30} & 2.35 & 2.36 & 2.36 \\
    \texttt{qwen2.5-coder (32B)} & 1.84 & 1.99 & 2.07 & 2.11 & 2.14 & 2.14 \\
    \texttt{codestral-2501 (22B)} & 1.44 & 1.55 & 1.62 & 1.69 & 1.73 & 1.75 \\
    \bottomrule
\end{tabular}
\vspace{-0.5cm}
\end{table}


Several observations emerge from these results. There is noticeable performance variability across models, but the top-performing models (\smalltt{gemini-2.0-flash}, \smalltt{gpt-4o}, \smalltt{gpt-o3-mini}) achieve relatively comparable speedups, particularly in the single-run scenario after sufficient iterations. Interestingly, reasoning models (\smalltt{gpt-o3-mini}, \smalltt{qwq}) did not consistently outperform top non-reasoning ones. While \smalltt{gpt-o3-mini} performed very strongly, \smalltt{qwq} was mid-pack. This suggests that while reasoning capabilities can be beneficial, the iterative feedback loop within \FrameworkName{} might provide sufficient guidance for capable non-reasoning models to perform well. Furthermore, models specialized for coding (\smalltt{qwen2.5-coder}, \smalltt{codestral}) did not demonstrate superior performance in this specific task; \smalltt{qwen2.5-coder} was average, and \smalltt{codestral} significantly lagged behind the others. This might indicate that proficiency in general code generation does not directly translate to effectively suggesting high-level transformations via a structured API.

Beyond final speedup, the models also differed significantly in their exploration efficiency, measured by the ratio of runnable schedules proposed. The performance differences correlate strongly with each model's ability to generate valid and legal schedules. As briefly noted in RQ3, the percentage of \textit{runnable} schedules varied significantly: \smalltt{gpt-o3-mini} achieved the highest rate ($\sim$40\%), aligning with its strong performance, while \smalltt{codestral} had the lowest ($\sim$15\%), consistent with its poor speedups. Models like \smalltt{gpt-4o} and \smalltt{llama3.3} also showed high runnable rates ($\sim$38\%), corresponding to their good performance. The full table of exploration efficiency for each LLM is provided in the appendix.

Furthermore, we experimented with several older-generation LLMs (\smalltt{CodeLlama}, \smalltt{CodeGemma}, \smalltt{DeepSeek-Coder-v2}) but excluded them from the main comparison. These models often failed to adhere to the structured output format or frequently hallucinated transformation commands, making programmatic interaction infeasible. This highlights that a baseline level of instruction following and structured output capability are essential for this approach.

In conclusion, while \FrameworkName{}'s methodology is applicable across various modern LLMs, the specific model choice impacts both the achievable performance and the efficiency of the optimization process. Top general-purpose models currently appear more suitable for this task than specialized coding models used in our study.
\subsubsection*{\textbf{RQ5: How does \FrameworkName{}'s performance compare to state-of-the-art polyhedral optimizer?}}
\label{subsubsec:RQ5} 

To better contextualize the performance gains reported in RQ1, we compare \FrameworkName{} against a recognized state-of-the-art polyhedral optimizer, Pluto~\cite{bondhugula2008practical}. 
Pluto employs a heuristic-driven approach to optimize for parallelism and data locality based on analyzing data dependencies within loop nests. Unlike \FrameworkName{}, Pluto does not rely on execution feedback during its decision process.

Overall, using the best-of-5 runs strategy, \BestOf{5}{30} achieves a geometric mean speedup of 2.94$\times$ over Pluto-optimized code across the 150 PolyBench instances (95\% CI: [2.88, 2.97]). \BestOf{5}{30} outperforms Pluto on 119 instances, matches it on 9 (within 5\%), and underperforms on 22.

\begin{figure*}[t]
    \centering
    
    \includegraphics[width=\linewidth]{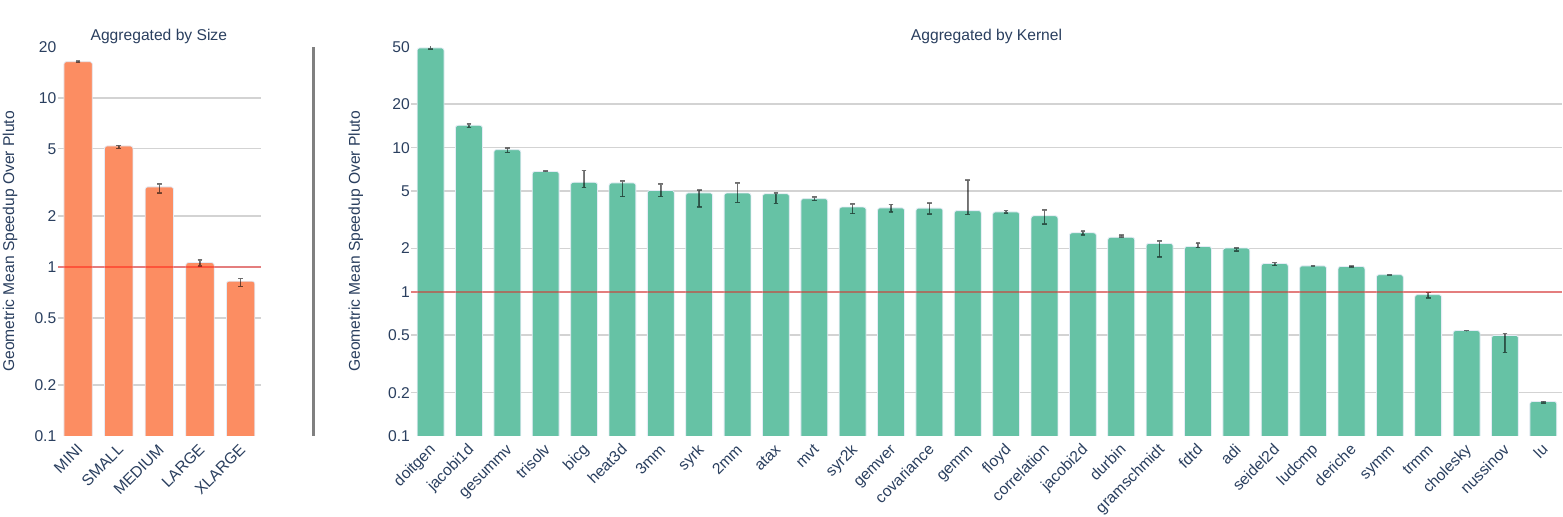}
    \vspace{-0.4cm}
    \caption{Geometric mean speedup of \BestOf{5}{30} compared to Pluto, aggregated by input size (left) and benchmark kernel (right). Error bars indicate 95\% CIs.}
    \vspace{-0.4cm}
    \label{fig:bar_charts_vs_pluto}
    
\end{figure*}

Figure~\ref{fig:bar_charts_vs_pluto} provides a breakdown of this relative performance, aggregated by input size and by benchmark kernel.
The aggregation by input size shows that \FrameworkName{}'s advantage over Pluto is most pronounced for smaller input sizes (16.35$\times$ for \smalltt{MINI}, 5.19$\times$ for \smalltt{SMALL}) and diminishes rapidly as size increases, becoming negligible for \smalltt{LARGE} (1.06$\times$) and even resulting in a slowdown for \smalltt{XLARGE} (0.82$\times$). This contrasts with the comparison against the original code (RQ1), where larger sizes saw greater speedups due to parallelization. Here, both \FrameworkName{} and Pluto implement parallelization, making the relative gains from parallelism less significant. Furthermore, Pluto applies size-agnostic heuristics, often tuned for large problem sizes. These heuristics can sometimes generate transformations detrimental to performance on smaller inputs. \FrameworkName{}, leveraging iterative feedback, appears better able to adapt its strategy or find schedules that avoid these pitfalls on smaller sizes.

Aggregated by kernel, \FrameworkName{} generally outperforms Pluto, though Pluto wins on a few kernels like \smalltt{lu}, \smalltt{nussinov}, and \smalltt{cholesky}. These cases can be justified by the fact that Pluto’s transformation space is much more sophisticated than the one that \FrameworkName{} can explore given the current set of primitives. Conversely, \FrameworkName{}'s significant advantage on many other kernels stems from its use of execution feedback, allowing it to optimize for measured performance rather than a proxy cost function, potentially finding schedules that Pluto's heuristic misjudges.

An interesting case is \smalltt{trisolv}. \FrameworkName{} achieves no speedup over the original code for this benchmark, yet shows a $\sim$6$\times$ geomean speedup over Pluto. This occurs because Pluto's schedule significantly slows down \smalltt{trisolv} compared to the original code, particularly for smaller/medium sizes, while \FrameworkName{} avoids such detrimental transformations.

To isolate the effect of these regressions, we compared \FrameworkName{} against a ``capped'' Pluto baseline where Pluto's performance is never worse than the original code (speedup $\ge 1\times$, simulating a scenario where Pluto's transformations are only applied if they don't cause a slowdown). Against this baseline, \BestOf{5}{30}'s overall geomean speedup drops from 2.94$\times$ to 1.78$\times$. The relative speedup on \smalltt{trisolv}, for instance, becomes 0.97$\times$. This confirms that a significant part of \FrameworkName{}'s advantage over Pluto stems from its ability to avoid the performance regressions that Pluto’s heuristic sometimes introduces. Nonetheless, \FrameworkName{} still maintains a considerable edge even after accounting for this.

Synthesizing these results clarifies why \FrameworkName{} can outperform a mature heuristic optimizer like Pluto. Whereas Pluto optimizes an analytical cost model as a proxy for performance, \FrameworkName{} optimizes measured performance, yielding two key advantages. First, it avoids performance regressions: guided by concrete feedback, the LLM quickly abandons slow paths (as seen on \smalltt{trisolv}), while Pluto may persist with a detrimental internal model. A large part of \FrameworkName{}’s gains come from steering clear of such pitfalls. Second, the iterative dialogue enables context-specific strategy adaptation. Whereas Pluto's ``one-size-fits-all" heuristics are often tuned for large problem sizes, \FrameworkName{} discovers specialized schedules through feedback-driven exploration. For instance, it learns to apply aggressive parallelization for large problems while favoring locality-enhancing transformations for smaller ones. 

\begin{figure}[t]
    \centering
    
    \includegraphics[width=0.9\linewidth]{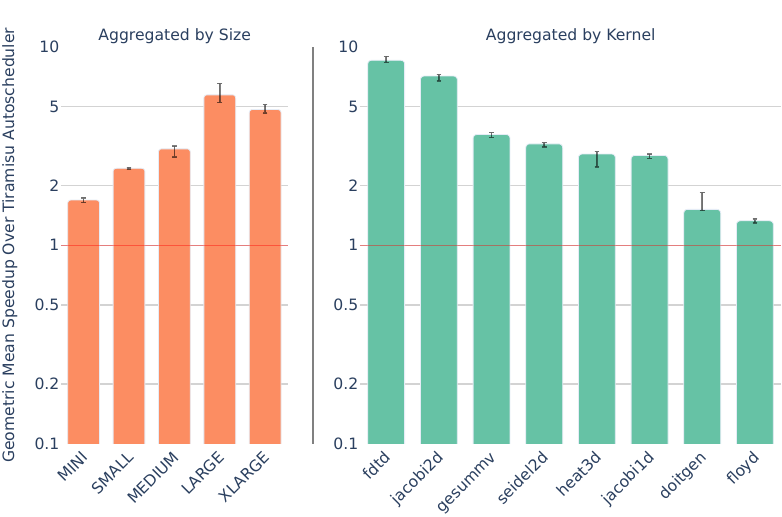}
    \vspace{-0.4cm}
    \caption{Geometric mean speedup of \BestOf{5}{30} compared to to the Tiramisu autoscheduler, aggregated by input size (left) and benchmark kernel (right) for the 8 supported benchmarks. Error bars indicate 95\% CIs.}
    \vspace{-0.4cm}
    \label{fig:bar_charts_vs_tiramisu}
    
\end{figure}

We also compare \FrameworkName{} against the deep-learning-based Tiramisu autoscheduler~\cite{baghdadi2021deep}, restricting evaluation to the 8 of 30 PolyBench benchmarks it supports. As noted in subsequent work~\cite{looper}, this autoscheduler cannot handle non-rectangular iteration domains or programs with multiple loop nests—limitations that exclude most of PolyBench. On these 8 benchmarks, \FrameworkName{} shows a clear advantage: \medianAt{30} attains a geometric mean speedup of 2.65$\times$ (95\% CI: [2.52, 2.69]) over the autoscheduler, rising to 3.23$\times$ (95\% CI: [3.14, 3.32]) with \BestOf{5}{30}. Figure~\ref{fig:bar_charts_vs_tiramisu} reports per-benchmark and per-size details.

This gap stems from three design differences: (1) \FrameworkName{} explores a richer transformation space, including skewing and reversal (critical for stencil dependencies in \smalltt{jacobi2d} and \smalltt{seidel2d}) that the autoscheduler does not support; (2) it employs a more flexible exploration strategy, iteratively combining and refining schedules rather than applying a fixed order, non-repetitive sequence; and (3) it is guided by ground-truth empirical feedback, while \cite{baghdadi2021deep} relies on predictions from an offline-trained cost model; this lets \FrameworkName{} adapt to the specific performance characteristics of the target hardware and avoid being misled by inaccurate model predictions.

\subsection{Ablation Study}

\subsubsection*{\textbf{RQ6: Is Feedback Important for Effective Optimization?}}
\label{subsubsec:RQ6}

\FrameworkName{}'s iterative process relies heavily on providing feedback (legality, validity, measured speedup/slowdown) to the LLM after each transformation attempt. This feedback loop serves a purpose analogous to Retrieval-Augmented Generation (RAG)~\cite{NEURIPS2020_6b493230}, providing external, factual context to ground the LLM's next suggestion. However, unlike typical RAG where context comes from a static corpus, \FrameworkName{}'s context is dynamically generated from interaction with the compiler and execution environment. Here, we assess the importance of this dynamic feedback mechanism.

To quantify its impact, we performed an ablation study comparing the standard \FrameworkName{} (``With Feedback") against a version where the LLM suggests transformations but receives no feedback on their outcome (``Without Feedback"). The LLM simply proposes schedules based on its initial analysis and internal models. Comparing the geometric mean speedups achieved with \smalltt{gemini-2.0-flash}, the ``With Feedback" configuration consistently outperforms the ``Without Feedback" version. Moreover, the performance gap widens significantly as the number of iterations ($T$) increases. At $T=30$, the standard \FrameworkName{} achieves approximately 23\% higher speedup in the single-run scenario (\medianAt{30}: 2.66$\times$ vs. 2.01$\times$) and 28\% higher speedup in the best-of-5 scenario compared to operating without feedback. This trend was even more pronounced when tested with \smalltt{gpt-4o}, where the single-run gap at $T=30$ reached approximately 40\% in favor of the feedback-driven approach.

The lack of feedback prevents the LLM from using its In-Context Learning capabilities to learn from its mistakes (e.g., avoiding patterns that lead to illegal schedules) or successes (e.g., refining schedules that yield speedups). Without guidance, the LLM essentially performs a blind, open-loop search, which is significantly less effective than the feedback-guided exploration in standard \FrameworkName{}. This clearly underscores the necessity of the interactive feedback loop for enabling the LLM to effectively navigate the complex transformation search space.

\subsubsection*{\textbf{RQ7: Does delegating code generation to the compiler improve LLM-based optimization?}}
\FrameworkName{}'s design intentionally separates concerns: the LLM provides high-level transformation guidance, while the compiler handles the complex tasks of applying transformations and ensuring legality via dependence analysis~\cite{feautrier_array_1988,violateddep}. To assess the benefits of this delegation, we evaluated an alternative approach where the LLM directly generates the transformed C code. 
We implemented a variant of the \FrameworkName{} workflow for this comparison. Instead of receiving transformation commands, this variant prompted the LLM to directly rewrite the input C kernel code extracted from PolyBench benchmarks. The goal was for the LLM to output optimized C code. The iterative dialogue structure remained, but the LLM's output and the subsequent verification steps were fundamentally different.

In this direct code-generation setup, correctness verification relied on output comparison: the generated code was compiled, executed, and its output compared against the original code's output for the default PolyBench input. If the outputs matched, the transformation was deemed ``\textit{provisionally legal}," and its execution time was measured. We acknowledge upfront that this method is imperfect and prone to false positives (i.e., incorrectly deeming an illegal transformation as legal, especially if the inputs used don't trigger the faulty behavior), unlike the formal dependence analysis used by Tiramisu in the standard \FrameworkName{} approach. The feedback provided to the LLM was simplified to reflect this process: reporting provisional legality failures (output mismatch), compilation errors, or successful execution with measured speedup.

Comparing the performance results, this direct code-generation approach consistently underperformed \FrameworkName{}. Measured against the original \medianAt{25} results, the direct code generation variant (after 25 iterations) yielded $\sim$14--16\% lower geometric mean speedups. This performance gap was observed consistently using two different LLMs, \smalltt{gemini-2.0-flash} and \smalltt{gpt-4o}. 

To quantify the risk of false positives inherent in the output comparison method, we performed a secondary check on the schedules deemed ``\textit{provisionally legal}" by the initial output comparison. Using randomly initialized input arrays (with a fixed seed), we re-ran the output comparison. Among the outperforming schedules generated (transformed codes that achieved a speedup $>1\times$), 17.6\% produced incorrect output in this second test, despite passing the initial check. Across all schedules explored that passed the initial check, 17.9\% were found to be illegal under random inputs. This high rate of correctness failures underscores the significant risk of using LLMs for direct code transformation without robust, formal verification. It contrasts with \FrameworkName{}'s design, where dependence analysis guarantees the legality of applied transformations.

Finally, the direct code-generation variant was substantially more expensive in terms of LLM interaction cost. We observed a $\sim$5.3$\times$ increase in token consumption compared to \FrameworkName{}. This is intuitively explained by the difference in output size: generating full C code requires significantly more tokens than outputting concise transformation commands.

In conclusion, these findings strongly support \FrameworkName{}'s strategy. Delegating the complexities of code generation and, crucially, formal legality verification to the compiler infrastructure leads to better optimization performance, ensures correctness, and significantly reduces the interaction cost compared to tasking the LLM with direct code generation for this problem.

\subsubsection*{\textbf{RQ8: How Does Providing Hardware Context in the Prompt Influence the Optimization Process?}}
\label{subsubsec:RQ8_HW_tuning}
A fundamental question in compiler optimization is the degree to which a strategy is tailored to a specific hardware target. To provide the LLM with this context, \FrameworkName{}'s system prompt explicitly details the target machine's hardware, including the CPU model, number of cores, and cache sizes. The goal is to allow the LLM to propose hardware-aware optimizations, such as selecting appropriate tile sizes.

To assess the impact of this information, we performed an ablation study where the optimization process was run without providing these hardware details in the prompt. Interestingly, the aggregate performance results showed no statistically significant difference compared to the standard configuration. We propose two complementary explanations for this observation.

First, the LLM's high-level strategy may be guided by general optimization principles learned from its training data—such as applying parallelism to large problems and tiling to improve locality—rather than by a nuanced interpretation of the specific hardware values provided. It is possible that current general-purpose LLMs are not yet proficient at translating hardware specifications into concrete decisions like choosing optimal tile sizes.

Second, the iterative feedback loop may provide a signal so strong that it overshadows any subtle strategy adjustments derived from the initial hardware description. The empirical performance measurements (speedup/slowdown) allow the LLM to learn to optimize for the hardware through trial and error, regardless of the initial description. In this scenario, the final performance converges even if the initial strategies differ.

Disentangling these effects would require a more fine-grained analysis, such as tracing the LLM's proposed transformation choices when presented with prompts containing varied hardware specifications. We leave this more detailed investigation for future work.

\subsection{Supplementary Analyses}

We conducted further analyses, detailed in the appendix, to characterize performance scaling and assess secondary methodological aspects. Regarding exploration effort (RQ9), we observed clear diminishing returns with both iterations per run ($T$) and the number of runs ($K$), supporting our choice of $T=30$ and $K=5$ for primary results. We also confirmed that prompting the LLM to continue exploring past premature termination attempts generally improved performance (RQ11). Finally, incorporating chain-of-thought prompting via initial program analysis and explicit reasoning steps provided tangible benefits (RQ10).






\section{Related Work}
LLMs have shown significant potential in software engineering tasks in recent years. Whether for code generation, code translation between programming languages, test case generation, or code summarization~\cite{zheng2023survey}. This has spurred significant interest in leveraging LLMs for compiler-related tasks, particularly in code transformation and optimization.

\subsection{Prior Work on LLM-Based Code Optimization}
\subsubsection{Domain-Specialized LLMs for Compiler Code Optimization.}

The Meta LLM Compiler~\cite{cummins2024meta} extends Code Llama~\cite{roziere2023code} with additional pretraining on LLVM-IR and assembly to specialize in low-level tasks (e.g., predicting optimization pass outcomes), followed by extensive fine-tuning for downstream compiler tasks such as flag selection for LLVM IR code-size reduction; it also introduces PassListEval to validate pass lists via unit testing and detect semantic breaks or crashes. Cummins et al.~\cite{cummins2023large} train a 7B-parameter LLVM-IR model to predict pass sequences for code-size reduction, relying on the compiler to apply transformations and ensure correctness. Grubisic et al.~\cite{grubisic2024compiler} add an iterative feedback loop that reports pass sequence validity and compilation errors, enabling the LLM to repair its outputs.

\subsubsection{High-Level Code Optimization}
Rosas et al.~\cite{rosas2024should} evaluate LLMs for OpenMP parallelization via direct code generation, validating correctness by checkpointing and comparing variable states against the original program. Shypula et al.~\cite{shypula2023learning} fine-tune on the Performance-Improving Edits (PIE) dataset curated from CodeNet~\cite{puri2021codenet}, assessing correctness with unit tests, while Duan et al.~\cite{PerfRL} fine-tune CodeT5~\cite{wang2021codet5} on PIE using reinforcement learning. LLM-Vectorizer~\cite{taneja2025llm} combines off-the-shelf GPT-4 with formal verification (Alive2~\cite{lopes2021alive2}) and a multi-agent finite-state machine to generate, validate, and repair vectorized code. Concurrently, CompilerGPT~\cite{compilergpt} employs an iterative LLM–compiler feedback loop in which the LLM analyzes Clang/GCC optimization reports and rewrites C++ to better trigger compiler heuristics, verifying correctness via a user-provided test harness.

\subsection{Positioning of Our Work}
Prior LLM-based code optimization largely uses domain-specialized models trained from scratch or fine-tuned on compiler-centric data (e.g., LLVM-IR, assembly) or on performance-improving edits, typically to directly generate optimized code. When framed as flag selection, the objective is usually code-size reduction rather than execution speed. Correctness is commonly checked via output comparison (which cannot guarantee semantic preservation) or via formal verification, whose support for complex optimizations and scalability to large programs is limited. In contrast, we leverage general-purpose, off-the-shelf LLMs (no fine-tuning) to iteratively select and guide optimizations. A key distinction from CompilerGPT~\cite{compilergpt}—which analyzes compiler reports and rewrites source to better trigger compiler heuristics—is that our LLM emits compiler API calls for loop transformations, while the compiler applies them and ensures legality via dependence analysis. This offloads code generation and provides correctness guarantees, unlike direct code-generation models and CompilerGPT, which rely on unit tests (whose potential for correctness failures we demonstrate in RQ7). Table~\ref{tab:related_works_comparaison} summarizes these differences.

In our implementation, the ability to perform source-level loop optimizations comes from the Tiramisu backend, which uses a polyhedral representation; \FrameworkName{} itself does not implement transformations. \FrameworkName{}’s contribution is the interaction-driven selection and sequencing of optimizations guided by compiler legality checks and measured performance. While the presented prototype manipulates source-level schedules, the paradigm is backend-agnostic: with GCC/Clang it could select flags or insert pragmas (e.g., unrolling, vectorization), and with LLVM it could orchestrate IR pass sequences. Thus, although our experiments focus on source-level loop transformations, \FrameworkName{} can drive any compiler-exposed optimization interface, regardless of abstraction level.

\begin{table}[t]
\caption{Comparison of LLM-Based Optimization Approaches.}
\vspace{-0.20cm}
\label{tab:related_works_comparaison}
\centering
\footnotesize
\setlength{\tabcolsep}{2.5pt}
\begin{tabular}{@{} l *{8}{c} @{}}
\toprule
\textbf{Feature} & \textbf{Ours} & \textbf{\cite{cummins2024meta}} & \textbf{\cite{cummins2023large}} & \textbf{\cite{grubisic2024compiler}} & \textbf{\cite{rosas2024should}} & \textbf{\cite{shypula2023learning}} & \textbf{\cite{PerfRL}} & \textbf{\cite{taneja2025llm}} \\
\cmidrule(r){1-1} \cmidrule(l){2-9}
Accelerating program execution      & \yes & \no  & \no  & \no  & \yes & \yes & \yes & \yes \\
Direct code generation              & \no  & \no  & \no  & \no  & \yes & \yes & \yes & \yes \\
Source-level code optimization      & \yes & \no  & \no  & \no  & \yes & \yes & \yes & \yes \\
Guarantees correctness              & \yes & \no  & \yes & \yes & \no  & \no  & \no  & \no  \\
Uses Chain-of-Thought               & \yes & \no  & \no  & \no  & \yes & \yes & \no  & \no  \\
Validity feedback                   & \yes & \no  & \no  & \yes & \no  & \no  & \no  & \yes \\
Refinement feedback                 & \yes & \no  & \no  & \no  & \no  & \no  & \no  & \no  \\
Uses Off-the-shelf LLM                & \yes & \no  & \no  & \no  & \yes & \no  & \no  & \yes \\
\bottomrule
\end{tabular}
\vspace{-0.45cm}
\end{table}

\subsection{Automatic Code Optimization Methods}
Classical approaches to automatic code optimization have evolved significantly, particularly within polyhedral and non-polyhedral compiler frameworks. Below, we outline key methodologies.

\subsubsection{Polyhedral Compilers and Autoschedulers}
The polyhedral model~\cite{Feautrier2011} is a mathematical model for representing code and code transformations and is used in state-of-the-art compilers to apply complex code transformations and reason about their correctness~\cite{wolf1991loop,lefebvre_automatic_1998,Qui00,thies_unified_2001,Darte_contraction_2005, baghdadi2015PhD,tiramisu,trifunovic_graphite_2010,polly,tobias_hexagonal_cgo13,Vasilache2018TensorCF,baghdadi2011speculation,merouani2020deep, pouchet.11.popl,baghdadi2020tiramisuDNNDenseSparse}.
Tools such as Pluto~\cite{bondhugula2008practical}, leverage the polyhedral model to represent loop nests enabling systematic exploration of loop transformations like tiling, fusion, and skewing. Pluto employs an integer linear programming (ILP) solver to derive affine transformations that optimize data locality and parallelism, guided by a linear objective function. PolyGym~\cite{brauckmann2021polygym} extends this paradigm with a reinforcement learning environment to explore schedules.

\subsubsection{Non-Polyhedral Autoschedulers}
Halide~\cite{adams2019learning} and TVM~\cite{chen2018tvm} adopt domain-specific auto-scheduling with deep-learning cost models. Halide combines beam search with a feedforward neural network to predict execution times from handcrafted features (e.g., loop nesting depth, memory access patterns). Similarly, AutoTVM~\cite{chen2018learning} uses gradient-based optimization to tune tensor operations for accelerators. These frameworks excel in domain-specific contexts (e.g., image processing, DNN kernels).

\section{Discussion and Future Work}
This work demonstrates that LLMs can effectively guide loop optimization through interactive dialogue, achieving speedups competitive with state-of-the-art compilers without fine-tuning. This compiler-LLM interaction-based approach presents a viable alternative to direct code generation or specialized model training. 

\FrameworkName{}'s effectiveness seems to hinge on the combination of the LLM's pattern recognition and reasoning with the compiler's rigorous analysis and execution feedback. The iterative loop, providing empirical performance data and legality checks, is crucial; removing it hampered optimization, confirming that the LLM learns dynamically from the interaction. Techniques encouraging structured reasoning, like initial analysis and explaining suggestions, also contributed positively.

However, the approach faces practical challenges. LLMs frequently proposed invalid or illegal transformations, leading to inefficient exploration and underscoring the necessity of the compiler's validation role. This inefficiency, coupled with the need for multiple runs to mitigate stochasticity and escape local optima, increased the time and computational cost. 

These limitations suggest clear directions for improvement. A key avenue is to enhance the agent's perception by enriching the feedback it receives from the environment. To enhance search efficiency, this feedback could detail the specific reasons for legality failures (e.g., the exact data dependency violated), helping the LLM learn correctness constraints more rapidly. To enable true hardware-specific tuning, the feedback could be augmented with data from hardware performance counters (HPCs). Instead of relying solely on wall-clock execution time, providing the LLM with metrics on cache miss rates or vector lane utilization for each attempted schedule would offer a much richer signal about the hardware-software interaction. This would allow the LLM to reason not just about whether a schedule is faster, but why it is faster. Beyond enhancing feedback, future work could also explore hybridizing LLM guidance with systematic search algorithms to better escape local optima, and developing dialogue summarization techniques to manage context length and reduce computational cost.  Despite the current limitations, using LLMs as interactive compiler assistants holds considerable promise for tackling complex performance optimization problems.

\section{Conclusion}
We investigated whether off-the-shelf Large Language Models can be effective at complex loop optimization. We introduced \FrameworkName{}, a framework that casts the LLM as an optimization agent interacting with a compiler environment. The agent acts by proposing transformations, which are then validated for legality and evaluated for performance by the compiler, feeding the results back as empirical feedback.

Our comprehensive evaluation on the PolyBench suite demonstrates the viability and effectiveness of this approach. \FrameworkName{} achieved a geometric mean speedup of 3.54x over original code and 2.94x over the Pluto optimizer using a best-of-5 strategy. This confirms that LLMs can successfully navigate challenging optimization spaces via iterative refinement based on empirical feedback, without needing specialized training. This work validates a novel paradigm for compiler optimization, showcasing the potential of leveraging general-purpose AI reasoning as a powerful component within automated performance engineering toolchains.

\section*{Appendix}
All supplementary materials, including extended analyses and results referenced throughout this paper, are available in the appendix. This appendix is included with our paper on arXiv, which can be found under the same title.

\section*{Acknowledgment}
This research has been partly supported by the Center for Artificial Intelligence and Robotics (CAIR) at New York University Abu Dhabi, funded by Tamkeen under the NYUAD Research Institute Award CG010.
The authors are grateful for the considerable computational resources provided by the Commit research group, led by Professor Saman Amarasinghe at the MIT Computer Science and Artificial Intelligence Laboratory (CSAIL). A portion of the research was also carried out on the High-Performance Computing resources at New York University Abu Dhabi.

\newpage

\bibliographystyle{IEEEtran}
\bibliography{sample-base}

@article{zheng2023survey,
  title={A survey of large language models for code: Evolution, benchmarking, and future trends},
  author={Zheng, Zibin and Ning, Kaiwen and Wang, Yanlin and Zhang, Jingwen and Zheng, Dewu and Ye, Mingxi and Chen, Jiachi},
  journal={arXiv preprint arXiv:2311.10372},
  year={2023}
}

@article{cummins2024meta,
  title={Meta large language model compiler: Foundation models of compiler optimization},
  author={Cummins, Chris and Seeker, Volker and Grubisic, Dejan and Roziere, Baptiste and Gehring, Jonas and Synnaeve, Gabriel and Leather, Hugh},
  journal={arXiv preprint arXiv:2407.02524},
  year={2024}
}

@article{cummins2023large,
  title={Large language models for compiler optimization},
  author={Cummins, Chris and Seeker, Volker and Grubisic, Dejan and Elhoushi, Mostafa and Liang, Youwei and Roziere, Baptiste and Gehring, Jonas and Gloeckle, Fabian and Hazelwood, Kim and Synnaeve, Gabriel and others},
  journal={arXiv preprint arXiv:2309.07062},
  year={2023}
}

@article{roziere2023code,
  title={Code llama: Open foundation models for code},
  author={Roziere, Baptiste and Gehring, Jonas and Gloeckle, Fabian and Sootla, Sten and Gat, Itai and Tan, Xiaoqing Ellen and Adi, Yossi and Liu, Jingyu and Sauvestre, Romain and Remez, Tal and others},
  journal={arXiv preprint arXiv:2308.12950},
  year={2023}
}

@article{grubisic2024compiler,
  title={Compiler generated feedback for large language models},
  author={Grubisic, Dejan and Cummins, Chris and Seeker, Volker and Leather, Hugh},
  journal={arXiv preprint arXiv:2403.14714},
  year={2024}
}

@article{rosas2024should,
  title={Should ai optimize your code? a comparative study of current large language models versus classical optimizing compilers},
  author={Rosas, Miguel Romero and Sanchez, Miguel Torres and Eigenmann, Rudolf},
  journal={arXiv preprint arXiv:2406.12146},
  year={2024}
}

@article{shypula2023learning,
  title={Learning performance-improving code edits},
  author={Shypula, Alexander and Madaan, Aman and Zeng, Yimeng and Alon, Uri and Gardner, Jacob and Hashemi, Milad and Neubig, Graham and Ranganathan, Parthasarathy and Bastani, Osbert and Yazdanbakhsh, Amir},
  journal={arXiv preprint arXiv:2302.07867},
  year={2023}
}

@article{PerfRL,
  title={Leveraging reinforcement learning and large language models for code optimization},
  author={Duan, Shukai and Kanakaris, Nikos and Xiao, Xiongye and Ping, Heng and Zhou, Chenyu and Ahmed, Nesreen K and Ma, Guixiang and Capota, Mihai and Willke, Theodore L and Nazarian, Shahin and others},
  journal={arXiv preprint arXiv:2312.05657},
  year={2023}
}

@inproceedings{taneja2025llm,
  title={Llm-vectorizer: Llm-based verified loop vectorizer},
  author={Taneja, Jubi and Laird, Avery and Yan, Cong and Musuvathi, Madan and Lahiri, Shuvendu K},
  booktitle={Proceedings of the 23rd ACM/IEEE International Symposium on Code Generation and Optimization},
  pages={137--149},
  year={2025}
}

@inproceedings{brauckmann2021polygym,
  title={Polygym: Polyhedral optimizations as an environment for reinforcement learning},
  author={Brauckmann, Alexander and Goens, Andr{\'e}s and Castrillon, Jeronimo},
  booktitle={2021 30th International Conference on Parallel Architectures and Compilation Techniques (PACT)},
  pages={17--29},
  year={2021},
  organization={IEEE}
}

@article{adams2019learning,
  title={Learning to optimize halide with tree search and random programs},
  author={Adams, Andrew and Ma, Karima and Anderson, Luke and Baghdadi, Riyadh and Li, Tzu-Mao and Gharbi, Micha{\"e}l and Steiner, Benoit and Johnson, Steven and Fatahalian, Kayvon and Durand, Fr{\'e}do and others},
  journal={ACM Transactions on Graphics (TOG)},
  volume={38},
  number={4},
  pages={1--12},
  year={2019},
  publisher={ACM New York, NY, USA}
}

@article{chen2018tvm,
  title={TVM: end-to-end optimization stack for deep learning},
  author={Chen, Tianqi and Moreau, Thierry and Jiang, Ziheng and Shen, Haichen and Yan, Eddie Q and Wang, Leyuan and Hu, Yuwei and Ceze, Luis and Guestrin, Carlos and Krishnamurthy, Arvind},
  journal={arXiv preprint arXiv:1802.04799},
  volume={11},
  number={2018},
  pages={20},
  year={2018},
  publisher={CoRR}
}

@article{chen2018learning,
  title={Learning to optimize tensor programs},
  author={Chen, Tianqi and Zheng, Lianmin and Yan, Eddie and Jiang, Ziheng and Moreau, Thierry and Ceze, Luis and Guestrin, Carlos and Krishnamurthy, Arvind},
  journal={Advances in Neural Information Processing Systems},
  volume={31},
  year={2018}
}

@article{puri2021codenet,
  title={Codenet: A large-scale ai for code dataset for learning a diversity of coding tasks},
  author={Puri, Ruchir and Kung, David S and Janssen, Geert and Zhang, Wei and Domeniconi, Giacomo and Zolotov, Vladimir and Dolby, Julian and Chen, Jie and Choudhury, Mihir and Decker, Lindsey and others},
  journal={arXiv preprint arXiv:2105.12655},
  year={2021}
}

@article{wang2021codet5,
  title={Codet5: Identifier-aware unified pre-trained encoder-decoder models for code understanding and generation},
  author={Wang, Yue and Wang, Weishi and Joty, Shafiq and Hoi, Steven CH},
  journal={arXiv preprint arXiv:2109.00859},
  year={2021}
}

@MISC{louis-noel_polybench_2010,
  author = {{Louis-Noel}, Pouchet},
  title = {{PolyBench} suite},
  howpublished = {http://www.cse.ohio-state.edu/{\textasciitilde}pouchet/software/polybench/},
  year = {2010},
  url = {http://www.cse.ohio-state.edu/~pouchet/software/polybench/}
}

@inproceedings{lopes2021alive2,
  title={Alive2: bounded translation validation for LLVM},
  author={Lopes, Nuno P and Lee, Juneyoung and Hur, Chung-Kil and Liu, Zhengyang and Regehr, John},
  booktitle={Proceedings of the 42nd ACM SIGPLAN International Conference on Programming Language Design and Implementation},
  pages={65--79},
  year={2021}
}

@inproceedings{tiramisu,
author = {Baghdadi, Riyadh and Ray, Jessica and Romdhane, Malek Ben and Del Sozzo, Emanuele and Akkas, Abdurrahman and Zhang, Yunming and Suriana, Patricia and Kamil, Shoaib and Amarasinghe, Saman},
title = {Tiramisu: a polyhedral compiler for expressing fast and portable code},
year = {2019},
isbn = {9781728114361},
publisher = {IEEE Press},
address = {New York, NY, USA},
booktitle = {Proceedings of the 2019 IEEE/ACM International Symposium on Code Generation and Optimization},
pages = {193–205},
numpages = {13},
location = {Washington, DC, USA},
series = {CGO 2019}
}

@article{bondhugula2008practical,
author = {Bondhugula, Uday and Hartono, Albert and Ramanujam, J. and Sadayappan, P.},
title = {A practical automatic polyhedral parallelizer and locality optimizer},
year = {2008},
issue_date = {June 2008},
publisher = {Association for Computing Machinery},
address = {New York, NY, USA},
volume = {43},
number = {6},
issn = {0362-1340},
url = {https://doi.org/10.1145/1379022.1375595},
doi = {10.1145/1379022.1375595},
journal = {SIGPLAN Not.},
month = jun,
pages = {101–113},
numpages = {13}
}

@inproceedings{NEURIPS2020_6b493230,
 author = {Lewis, Patrick and Perez, Ethan and Piktus, Aleksandra and Petroni, Fabio and Karpukhin, Vladimir and Goyal, Naman and K\"{u}ttler, Heinrich and Lewis, Mike and Yih, Wen-tau and Rockt\"{a}schel, Tim and Riedel, Sebastian and Kiela, Douwe},
 booktitle = {Advances in Neural Information Processing Systems},
 editor = {H. Larochelle and M. Ranzato and R. Hadsell and M.F. Balcan and H. Lin},
 pages = {9459--9474},
 publisher = {Curran Associates, Inc.},
 title = {Retrieval-Augmented Generation for Knowledge-Intensive NLP Tasks},
 url = {https://proceedings.neurips.cc/paper_files/paper/2020/file/6b493230205f780e1bc26945df7481e5-Paper.pdf},
 volume = {33},
 year = {2020}
}

@inproceedings{NEURIPS2022_9d560961,
 author = {Wei, Jason and Wang, Xuezhi and Schuurmans, Dale and Bosma, Maarten and ichter, brian and Xia, Fei and Chi, Ed and Le, Quoc V and Zhou, Denny},
 booktitle = {Advances in Neural Information Processing Systems},
 editor = {S. Koyejo and S. Mohamed and A. Agarwal and D. Belgrave and K. Cho and A. Oh},
 pages = {24824--24837},
 publisher = {Curran Associates, Inc.},
 title = {Chain-of-Thought Prompting Elicits Reasoning in Large Language Models},
 url = {https://proceedings.neurips.cc/paper_files/paper/2022/file/9d5609613524ecf4f15af0f7b31abca4-Paper-Conference.pdf},
 volume = {35},
 year = {2022}
}

@inproceedings{NEURIPS2020_1457c0d6,
 author = {Brown, Tom and Mann, Benjamin and Ryder, Nick and Subbiah, Melanie and Kaplan, Jared D and Dhariwal, Prafulla and Neelakantan, Arvind and Shyam, Pranav and Sastry, Girish and Askell, Amanda and Agarwal, Sandhini and Herbert-Voss, Ariel and Krueger, Gretchen and Henighan, Tom and Child, Rewon and Ramesh, Aditya and Ziegler, Daniel and Wu, Jeffrey and Winter, Clemens and Hesse, Chris and Chen, Mark and Sigler, Eric and Litwin, Mateusz and Gray, Scott and Chess, Benjamin and Clark, Jack and Berner, Christopher and McCandlish, Sam and Radford, Alec and Sutskever, Ilya and Amodei, Dario},
 booktitle = {Advances in Neural Information Processing Systems},
 editor = {H. Larochelle and M. Ranzato and R. Hadsell and M.F. Balcan and H. Lin},
 pages = {1877--1901},
 publisher = {Curran Associates, Inc.},
 title = {Language Models are Few-Shot Learners},
 url = {https://proceedings.neurips.cc/paper_files/paper/2020/file/1457c0d6bfcb4967418bfb8ac142f64a-Paper.pdf},
 volume = {33},
 year = {2020}
}

@inproceedings{violateddep,
author = {Vasilache, Nicolas and Bastoul, Cedric and Cohen, Albert and Girbal, Sylvain},
title = {Violated Dependence Analysis},
year = {2006},
isbn = {1595932828},
publisher = {Association for Computing Machinery},
address = {New York, NY, USA},
url = {https://doi.org/10.1145/1183401.1183448},
doi = {10.1145/1183401.1183448},
booktitle = {Proceedings of the 20th Annual International Conference on Supercomputing},
pages = {335–344},
numpages = {10},
location = {Cairns, Queensland, Australia},
series = {ICS '06}
}

@Inbook{Feautrier2011,
author="Feautrier, Paul
and Lengauer, Christian",
editor="Padua, David",
title="Polyhedron Model",
bookTitle="Encyclopedia of Parallel Computing",
year="2011",
publisher="Springer US",
address="Boston, MA",
pages="1581--1592",
isbn="978-0-387-09766-4",
doi="10.1007/978-0-387-09766-4_502",
url="https://doi.org/10.1007/978-0-387-09766-4_502"
}

@inproceedings{lattner2004llvm,
  title={LLVM: A compilation framework for lifelong program analysis \& transformation},
  author={Lattner, Chris and Adve, Vikram},
  booktitle={International symposium on code generation and optimization, 2004. CGO 2004.},
  pages={75--86},
  year={2004},
  organization={IEEE}
}

@article{stallman2002gnu,
  title={GNU compiler collection internals},
  author={Stallman, Richard M},
  journal={Free Software Foundation},
  volume={46},
  year={2002}
}

@inproceedings{ansel2014opentuner,
  title={Opentuner: An extensible framework for program autotuning},
  author={Ansel, Jason and Kamil, Shoaib and Veeramachaneni, Kalyan and Ragan-Kelley, Jonathan and Bosboom, Jeffrey and O'Reilly, Una-May and Amarasinghe, Saman},
  booktitle={Proceedings of the 23rd international conference on Parallel architectures and compilation},
  pages={303--316},
  year={2014}
}

@ARTICLE{8423171,
  author={Balaprakash, Prasanna and Dongarra, Jack and Gamblin, Todd and Hall, Mary and Hollingsworth, Jeffrey K. and Norris, Boyana and Vuduc, Richard},
  journal={Proceedings of the IEEE}, 
  title={Autotuning in High-Performance Computing Applications}, 
  year={2018},
  volume={106},
  number={11},
  pages={2068-2083},
  doi={10.1109/JPROC.2018.2841200}}

@Inbook{Vuduc2011,
author="Vuduc, Richard W.",
editor="Padua, David",
title="Autotuning",
bookTitle="Encyclopedia of Parallel Computing",
year="2011",
publisher="Springer US",
address="Boston, MA",
pages="102--105",
isbn="978-0-387-09766-4",
doi="10.1007/978-0-387-09766-4_68",
url="https://doi.org/10.1007/978-0-387-09766-4_68"
}

@misc{compilergpt,
      title={CompilerGPT: Leveraging Large Language Models for Analyzing and Acting on Compiler Optimization Reports}, 
      author={Peter Pirkelbauer and Chunhua Liao},
      year={2025},
      eprint={2506.06227},
      archivePrefix={arXiv},
      primaryClass={cs.PL},
      url={https://arxiv.org/abs/2506.06227}, 
}

@article{looper,
      title={LOOPer: A Learned Automatic Code Optimizer For Polyhedral Compilers}, 
      author={Massinissa Merouani and Khaled Afif Boudaoud and Iheb Nassim Aouadj and Nassim Tchoulak and Islem Kara Bernou and Hamza Benyamina and Fatima Benbouzid-Si Tayeb and Karima Benatchba and Hugh Leather and Riyadh Baghdadi},
      year={2025},
      eprint={2403.11522},
      archivePrefix={arXiv},
      primaryClass={cs.PL},
        journal={arXiv preprint arXiv:2403.11522},
      url={https://arxiv.org/abs/2403.11522}, 
}

@article{baghdadi2021deep,
  title={A deep learning based cost model for automatic code optimization},
  author={Baghdadi, Riyadh and Merouani, Massinissa and Leghettas, Mohamed-Hicham and Abdous, Kamel and Arbaoui, Taha and Benatchba, Karima and others},
  journal={Proceedings of Machine Learning and Systems},
  volume={3},
  pages={181--193},
  year={2021}
}

@misc{baghdadi2020tiramisuDNNDenseSparse,
title={TIRAMISU: A Polyhedral Compiler for Dense and Sparse Deep Learning},
author={Riyadh Baghdadi and Abdelkader Nadir Debbagh and Kamel Abdous and Fatima Zohra Benhamida and Alex Renda and Jonathan Elliott Frankle and Michael Carbin and Saman Amarasinghe},
year={2020},
eprint={2005.04091},
archivePrefix={arXiv},
primaryClass={cs.DC}
}

@misc{baghdadi2011speculation,
title={The Potential of Synergistic Static, Dynamic and Speculative Loop Nest Optimizations for Automatic Parallelization},
author={Riyadh Baghdadi and Albert Cohen and Cedric Bastoul and Louis-Noel Pouchet and Lawrence Rauchwerger},
year={2011},
eprint={1111.6756},
archivePrefix={arXiv},
primaryClass={cs.DC}
}

@phdthesis{baghdadi2015PhD,
title={Improving tiling, reducing compilation time, and extending the scope of polyhedral compilation},
author={Baghdadi, Riyadh},
year={2015},
school={Paris 6}
}

@mastersthesis{merouani2020deep,
  title={A deep learning based cost model for automatic code optimization in tiramisu},
  author={Merouani, Massinissa and Leghettas, Mohamed-Hicham and Baghdadi, Riyadh and Arbaoui, Taha and Benatchba, Karima},
  year={2020},
  school={ESI}
}

@article{wolf1991loop,
  title={A loop transformation theory and an algorithm to maximize parallelism},
  author={Wolf, Michael E and Lam, Monica S},
  journal={IEEE transactions on parallel and distributed systems},
  volume={2},
  number={4},
  pages={452--471},
  year={1991},
  publisher={IEEE}
}

@MISC{trifunovic_graphite_2010,
  author = {Trifunovic, Konrad and Cohen, Albert and Edelsohn, David and Li,
	Feng and Grosser, Tobias and Jagasia, Harsha and Ladelsky, Razya
	and Pop, Sebastian and Sjodin, Jan and Upadrasta, Ramakrishna},
  title = {{GRAPHITE} Two Years After: First Lessons Learned From {Real-World}
	Polyhedral Compilation},
  month = jan,
  year = {2010},
  shorttitle = {{GRAPHITE} Two Years After}
}

@article{polly,
  author = {Grosser, Tobias and Groslinger, Armin and Lengauer, Christian},
  ee = {http://dx.doi.org/10.1142/S0129626412500107},
  journal = {Parallel Processing Letters},
  number = 4,
  title = {Polly - Performing Polyhedral Optimizations on a Low-Level Intermediate Representation.},
  url = {http://dblp.uni-trier.de/db/journals/ppl/ppl22.html#GrosserGL12},
  volume = 22,
  year = 2012
}

@article{Vasilache2018TensorCF,
  title={Tensor Comprehensions: Framework-Agnostic High-Performance Machine Learning Abstractions},
  author={Nicolas Vasilache and Oleksandr Zinenko and Theodoros Theodoridis and Priya Goyal and Zach DeVito and William S. Moses and Sven Verdoolaege and Andrew Adams and Albert Cohen},
  journal={CoRR},
  year={2018},
  volume={abs/1802.04730}
}

@InProceedings{pouchet.11.popl,
  author = 	 { Louis-No{\"e}l Pouchet and Uday Bondhugula and
                  C{\'e}dric Bastoul and Albert Cohen and J. Ramanujam and
                  P. Sadayappan and Nicolas Vasilache },
  title = 	 { Loop Transformations: Convexity, Pruning and Optimization },
  booktitle = { 38th ACM SIGACT-SIGPLAN Symposium on Principles of Programming Languages (POPL'11)},
  address = { Austin, TX},
  month = 	 jan,
  year = 	 {2011},
  pages = {549--562},
  publisher =	 { ACM Press }
}

@inproceedings{tobias_hexagonal_cgo13,
 author = {Grosser, Tobias and Cohen, Albert and Holewinski, Justin and Sadayappan, P. and Verdoolaege, Sven},
 title = {Hybrid Hexagonal/Classical Tiling for GPUs},
 booktitle = {Proceedings of Annual IEEE/ACM International Symposium on Code Generation and Optimization},
 series = {CGO '14},
 year = {2014},
 location = {Orlando, FL, USA},
 pages = {66:66--66:75},
 articleno = {66},
 numpages = {10},
 acmid = {2544160},
 publisher = {ACM},
 address = {New York, NY, USA},
}

@INPROCEEDINGS{feautrier_array_1988,
  author = {Feautrier, P.},
  title = {Array expansion},
  booktitle = {Proceedings of the 2nd international conference on Supercomputing},
  year = {1988},
  pages = {429--441},
  address = {St. Malo, France},
  publisher = {{ACM}},
  doi = {10.1145/55364.55406},
  isbn = {0-89791-272-1},
  url = {http://portal.acm.org/citation.cfm?id=55406}
}

@article{Darte_contraction_2005,
 author = {Darte, Alain and Huard, Guillaume},
 title = {New Complexity Results on Array Contraction and Related Problems},
 journal = {J. VLSI Signal Process. Syst.},
 issue_date = {May       2005},
 volume = {40},
 number = {1},
 month = may,
 year = {2005},
 issn = {0922-5773},
 pages = {35--55},
 numpages = {21},
 url = {http://dx.doi.org/10.1007/s11265-005-4937-3},
 doi = {10.1007/s11265-005-4937-3},
 acmid = {1050593},
 publisher = {Kluwer Academic Publishers},
 address = {Hingham, MA, USA},
}

@Article{Qui00,
  author = 	 {F. Quiller\'e and S. Rajopadhye},
  title = 	 {Optimizing Memory Usage in the Polyhedral Model},
  journal = 	 {ACM Trans. on Programming Languages and Systems},
  year = 	 2000,
  volume =	 22,
  number =	 5,
  pages =	 {773--815},
  month =	 sep
}

@ARTICLE{lefebvre_automatic_1998,
  author = {Lefebvre, Vincent and Feautrier, Paul},
  title = {Automatic storage management for parallel programs},
  journal = {Parallel Computing},
  year = {1998},
  volume = {24},
  pages = {649--671},
  doi = {10.1016/S0167-8191(98)00029-5},
  issn = {01678191}
}

@INPROCEEDINGS{thies_unified_2001,
  author = {Thies, William and Vivien, Fr\'{e}d\'{e}ric and Sheldon, Jeffrey
	and Amarasinghe, Saman},
  title = {A unified framework for schedule and storage optimization},
  year = {2001},
  booktitle = 	{Proc.\ of the 2001 {PLDI} Conf.},
}

\clearpage
\appendices

\section{System prompt}
\label{app:system_prompt}

\definecolor{customBorder}{HTML}{66C2A5}
\begin{figure*}[!h]
\centering
\lstset{
  basicstyle=\ttfamily\tiny,
  breaklines=true,
  frame=single,
  rulecolor=\color{customBorder},
  backgroundcolor=\color{gray!10},
  columns=fullflexible,
  framerule=1pt,
  xleftmargin=0.5em,
  xrightmargin=0.5em,
  framexleftmargin=0.5em
}
\begin{lstlisting}
You are a compiler optimization assistant. Your task is to iteratively explore and suggest sequences of loop transformations (i.e. a schedule) for a given C++ loop nest to minimize its execution time. You will interact iteratively with a compiler that uses the Tiramisu API.

# Overview:
Initially, the compiler will show you a loop nest and ask you to analyze it. After you provide an analysis, the compiler will ask you to start the iterative transformations exploration. You will suggest loop transformations that the compiler should try. The compiler will apply the suggested transformations using the Tiramisu API and let you know whether the transformations are legal or not.If the transformations are legal, the compiler will execute the transformed program and report the speedup compared to the original execution time of the program before transformations. This process continues until you indicate there are no more interesting transformations to try.

# Input Format:
The compiler will present the C++ loop nest to be optimized. The input loop nest will be annotated with comments to give an ID to each computation block. The comment will have this structure `// comp_ID: <string>` e.g. `// comp_ID: comp05`. You need these IDs for specifying where to apply each transformation (as explained later). The compiler will also provide the initial execution time of the program before any transformations are applied.

# Analysis Phase:
Before starting the optimization process, the compiler will first ask you to analyze the loop nest. At this stage, focus only on analyzing the input program, do not suggest transformations until prompted. You may structure your analysis as you see fit, but it should provide insights into the structure of the loop nest, the computations being performed, and the program as a whole

# Schedule Suggestions:
To form a transformation suggestion, use the transformation commands listed below. Use the comp_IDs along with loop levels to specify where to apply transformations. A loop level should be specified by the letter `L` followed by the depth level of the loop in question. For example, to parallelize the outermost loop of comp05, you should say `comp05.Parallelize(L0)`, and to parallelize the second loop, you should say `comp05.Parallelize(L1)`.

You may suggest one or more transformations at a time. You can combine multiple transformations (forming a schedule) by joining the transformation commands with a `+` sign. For example `comp12.Parallelize(L0)+comp35.Unroll(L3,16)`.

If you have no more suggestions, use the no_further_transformations command to indicate to the compiler that no further transformations are planned. The compiler may ask you to explore more if it deems your suggestions insufficient. 
  
For the compiler to parse your response, format your suggestions as follows, replacing the comment with appropriate content:

```
Reasoning:
// Here you should insert your rationale for the new list of transformations and discuss the result of your previous suggestion based on the compiler's feedback.

New full list of transformations:
<schedule> /* Insert your new suggested sequence of transformations here */ </schedule>
```

Your full suggested sequence of transformations should be placed between the <schedule> and </schedule> tags in a single line, using the format explained earlier. For example: 

`<schedule>comp12.Parallelize(L0)+comp12.Tile2D(L1,L2,128,128)+comp35.Unroll(L3,16)</schedule>`

You can revoke transformations, modify them, extend them, or reorder them as necessary. Feel free to explore as many suggestions as you wish. There is no limit on the number of iterations.

# Supported Transformation Commands:
Below is the syntax of each supported transformation command:

- Loop Fusion: `<comp_ID_1>.Fuse(<comp_ID_2>, L<level>)`
- Loop Interchange: `<comp_ID>.Interchange(L<level1>, L<level2>)`
- Loop Parallelization: `<comp_ID>.Parallelize(L<level>)`
- 2D Loop Tiling: `<comp_ID>.Tile2D(L<level1>, L<level2>, <tiling_factor1>, <tiling_factor2>)`
- 3D Loop Tiling: `<comp_ID>.Tile3D(L<level1>, L<level2>, L<level3>, <tiling_factor1>, <tiling_factor2>, <tiling_factor3>)`
- Loop Unrolling: `<comp_ID>.Unroll(L<level>, <unrolling_factor>)`
- Loop Skewing: `<comp_ID>.Skew(L<level1>, L<level2>)`
- Loop Reversal: `<comp_ID>.Reverse(L<level>)`

# Benchmarking Setup:
Your suggested transformations will be applied using the Tiramisu compiler. Execution time will be measured on a machine equipped with the following CPU:

Model name: Intel(R) Xeon(R) CPU E5-2695 v2 @ 2.40GHz  
Thread(s) per core: 2  
Core(s) per socket: 12  
Socket(s): 2  
CPU(s): 48  
CPU max MHz: 3200.0000  
CPU min MHz: 1200.0000  
Caches (sum of all):  
    L1d: 768 KiB (24 instances)  
    L1i: 768 KiB (24 instances)  
    L2: 6 MiB (24 instances)  
    L3: 60 MiB (2 instances)

# General Notes:
- To unroll an innermost loop, you can use `L-1` as the loop level selector.
- Regarding loop skewing, the compiler will automatically determine the appropriate skewing factors by running a solver. Loop skewing can either enable parallelization (of one of the two skewed loops) or improve locality. Skewing works only if applied on a pair of perfectly nested loops.
- If a compound transformation is illegal or crashes, consider revoking some components of the schedule to identify the cause.
- Consider the following potential fixes for crashes:
    - If the sequence of transformations involves unrollings, you may consider reordering your list of transformations so that unrollings appear at the end.
    - If the sequence of transformations involves fusion, you may consider reordering your list of transformations so that fusion appears at the beginning.  Also, keep in mind that after applying fusion at level X, the two used comp_IDs will point to the same fused block up to level X. For example, comp04.fuse(comp05, L2)+comp04.Parallelize(L1) is strictly equivalent to comp04.fuse(comp05, L2)+comp05.Parallelize(L1) since comp04 at L1 and comp05 at L1 point to the same loop post-fusion. So ensure you are not applying the same transformation with different comp_IDs.

\end{lstlisting}
\caption{System prompt used in \FrameworkName{}.}
\label{fig:COLLM_system_prompt}

\end{figure*}

The full system prompt used for the LLM during the optimization process is shown in Figure~\ref{fig:COLLM_system_prompt}.

\section{Bootstrapping procedure for confidence interval estimation}
\label{app:bootstrapping_details}
To construct a confidence interval for the geometric mean of median speedups, we utilized a bootstrap resampling approach. For each of the 1000 bootstrap iterations, we resampled with replacement from the 40 speedup measurements for each of the 150 benchmark programs. For each bootstrap sample, we first computed the median speedup for each benchmark across the resampled runs. We then calculated the geometric mean of these 150 median speedups. This process resulted in a distribution of 1000 bootstrap geometric mean of median values. We derived a 95\% confidence interval by taking the 2.5th and 97.5th percentiles of this bootstrap distribution. Bootstrapping was chosen as a suitable method due to the stochastic nature of our optimization algorithm and the complexity of the 'geometric mean of medians' statistic, for which analytical confidence interval formulas are not readily available. Furthermore, bootstrapping is a non-parametric method, making no assumptions about the underlying distribution of speedup values, and is robust to potential outliers in the data. This approach provides a data-driven and statistically sound method to estimate the uncertainty associated with our performance metric and to assess its stability across different algorithm executions.


\section{Extended Results and Analysis}
\label{app:extended_results}

\subsubsection*{\textbf{RQ9: How does performance scale with the number of runs and iterations?}}
\label{subsubsec:RQ9}

\FrameworkName{}'s performance is influenced by the exploration depth (number of iterations $T$ per run) and breadth (number of runs $K$ in best-of-$K$ scenarios). We analyze how the typical speedup scales with these parameters.

\begin{figure}[h]
    \centering
    \includegraphics[width=\linewidth]{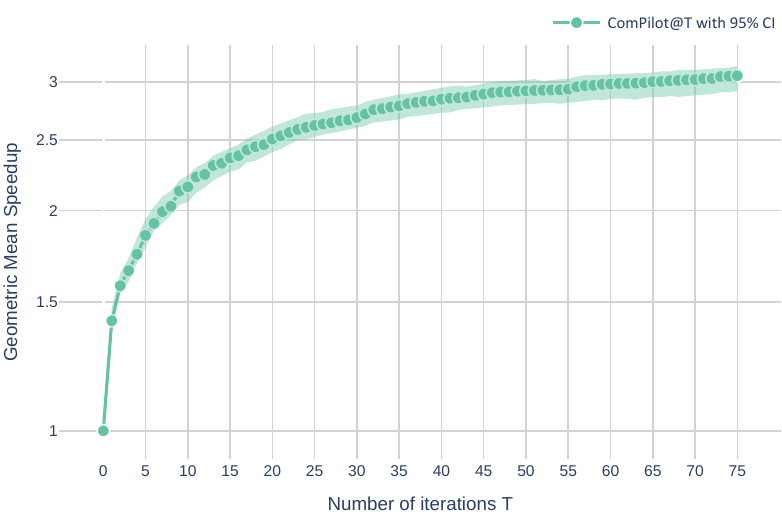}
    \caption{\medianAt{T} (across the entire benchmarks set) as a function of the number of iterations ($T$) per run.}
    \label{fig:speedup_vs_iterations}
    
\end{figure}

First, we examine the single-run speedup \medianAt{T} as iterations $T$ increase from 1 to 75 (Figure~\ref{fig:speedup_vs_iterations}). The results show clear diminishing returns with respect to the number of iterations. Speedup climbs rapidly initially (1.41$\times$ at $T=1$, 2.15$\times$ at $T=10$) but slows considerably later (2.68$\times$ at $T=30$, reaching only 3.06$\times$ at $T=75$). This saturation suggests that while longer dialogues allow for refinement and discovery of further optimizations, the most impactful transformations within the capabilities of the current setup are often found within the first few dozen iterations. The achievable performance is likely bounded by the expressiveness of the available transformation primitives and the inherent complexity of optimizing certain loop structures. Based on this curve, we selected $T=30$ iterations for reporting our primary results as it captures most of the gains efficiently.

\begin{figure}[h]
    \centering
    \includegraphics[width=\linewidth]{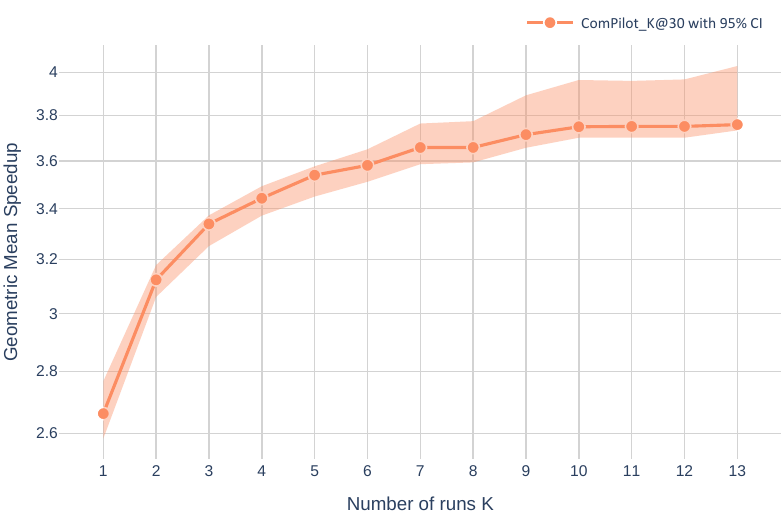}
    \caption{\BestOf{K}{30} (across the entire benchmarks set) as a function of the number of runs ($K$).}
    \label{fig:speedup_vs_runs}
    
\end{figure}

Next, fixing iterations at $T=30$, we study how the typical best-of-$K$ speedup (\BestOf{K}{30}) scales as the number of runs $K$ increases from 1 to 13 (Figure~\ref{fig:speedup_vs_runs}).
Similar diminishing returns are observed. Moving from one run ($K=1$, 2.66$\times$) to five runs ($K=5$, 3.54$\times$) offers a substantial boost by exploring diverse optimization paths. However, gains taper off afterward ($K=10$ yields 3.75$\times$, $K=13$ yields 3.82$\times$). While multiple runs effectively exploit the LLM's stochastic nature to explore diverse optimization avenues, the search spaces covered by different runs have considerable overlap, and the performance remains ultimately constrained by the optimization potential within \FrameworkName{}’s search space for these benchmarks. We selected $K=5$ as our representative multi-run scenario.

\begin{figure}[h]
    \centering
    \includegraphics[width=\linewidth]{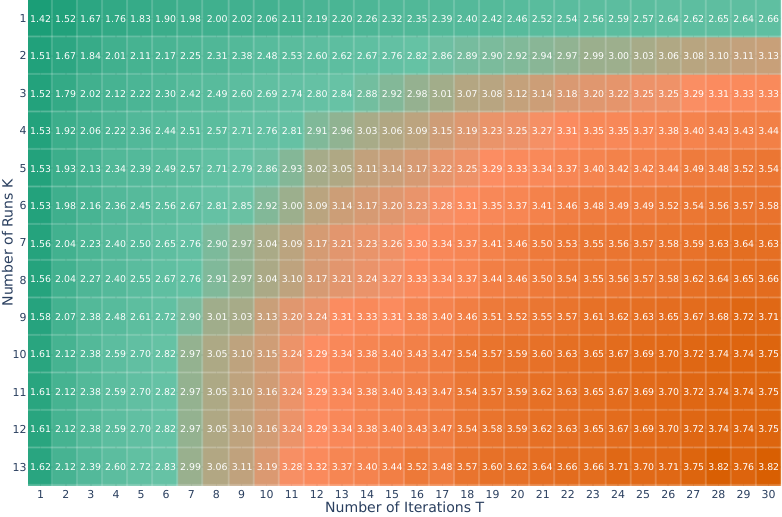} 
    \caption{Heatmap visualizing \BestOf{K}{T} as a function of both the number of iterations ($T$, x-axis) and the number of runs ($K$, y-axis).}
    \label{fig:speedup_heatmap}
    
\end{figure}

Figure~\ref{fig:speedup_heatmap} provides a heatmap of \BestOf{K}{T} over the $T$ (1-30) and $K$ (1-13) grid, visualizing the interplay between these parameters. The heatmap visually confirms the diminishing returns in both dimensions, showing the most significant speedup improvements occur at lower $T$ and $K$ values. It also illustrates the trade-off between investing in longer single runs versus performing multiple shorter runs to reach a given performance level.

\subsubsection*{\textbf{RQ10: How Important is Chain-of-Thought Reasoning in this context?}}
\label{subsubsec:RQ10} 

\FrameworkName{} incorporates two mechanisms akin to Chain-of-Thought (CoT)~[29] to potentially improve the LLM's reasoning: prompting for an initial program analysis before optimization begins, and requiring explicit reasoning before each schedule suggestion. These aim to encourage structured thinking. We evaluate their impact through ablation studies.

First, we removed the initial program analysis step. Compared to the standard \FrameworkName{} (using \smalltt{gemini-2.0-flash}), this resulted in consistently lower performance across iterations. At $T=30$, the geomean speedup dropped by \textasciitilde 8\% for the single-run scenario (\medianAt{30}: 2.42$\times$ vs. 2.66$\times$) and by \textasciitilde 4\% for the best-of-5 scenario. A similar trend, with an even larger gap (\textasciitilde 14\% for single-run), was observed using \smalltt{GPT-4o}. This suggests the upfront analysis provides a tangible benefit to the optimization process.

Second, we removed the requirement for the LLM to provide reasoning, forcing the LLM to output only the \smalltt{<schedule>} tag. With \smalltt{GPT-4o}, this consistently reduced single-run (\medianAt{T}) performance by \textasciitilde 11\%. Interestingly, with \smalltt{gemini-2.0-flash}, single-run performance was largely unaffected. However, for the best-of-5 scenario (\BestOf{5}{T}), omitting reasoning led to a 4-7\% performance drop for both LLMs.

Overall, both CoT-inspired components appear beneficial, although their impact varies. The initial program analysis provides a consistent, measurable performance boost. The utility of per-iteration reasoning seems more pronounced for certain models (like \smalltt{GPT-4o}) or potentially surfaces more strongly when leveraging multiple runs (best-of-$K$ scenario). Incorporating these structured reasoning steps generally contributes positively to \FrameworkName{}'s effectiveness.

\subsubsection*{\textbf{RQ11: Importance of pushing the LLM for more exploration.}}
\begin{figure}[h]
    \centering
    \includegraphics[width=\linewidth]{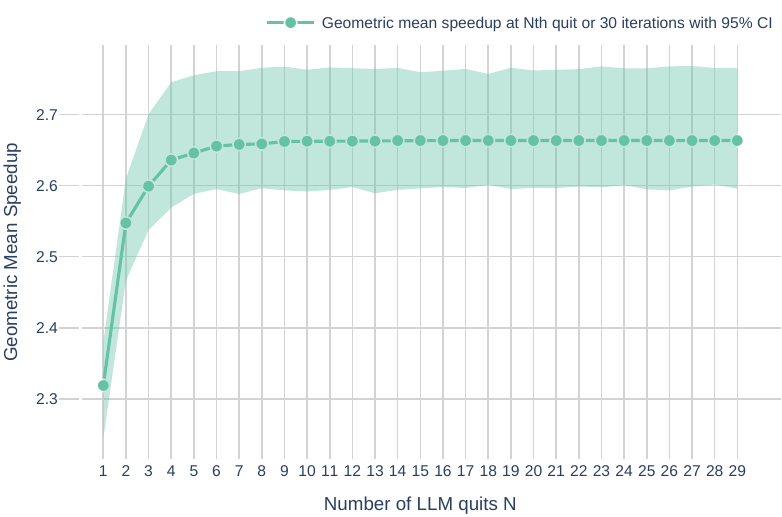}
    \caption{Geometric mean single-run speedup at Nth LLM quit or at iteration $T=30$, whichever occurs first, across all 150 instances as a function of the number of quits ($N$). Error bars represent 95\% bootstrap confidence intervals.}
    \label{fig:speedup_vs_LLM_quits}
    
\end{figure}

As described in Section~2, the LLM often terminates the optimization process prematurely once it achieves substantial speedup improvements or encounters multiple unsuccessful optimization attempts. To highlight the importance of pushing forward the LLM to further exploration, we plot in Figure~\ref{fig:speedup_vs_LLM_quits} the geometric mean of median speedups per benchmark across 40 runs, with each run stopping either at the LLM’s N‑th quit attempt or at the maximum iteration count $T=30$, whichever occurs first.

The results indicate that prompting the LLM to continue exploring leads to improved performance at the initial quitting attempts. Notably, a comparable speedup to \medianAt{30} is only achieved after the LLM is pushed to explore beyond its fifth quit attempt. However, additional pushing beyond this point yields diminishing returns, as the LLM tends to terminate more frequently after only a few unsuccessful iterations (\textasciitilde1 iteration) or marginal speedup gains. This latest behavior is only observed in less than ~20\% of the time. We also observe that in 2\% of cases, the LLM insists on terminating the conversation when prompted to continue exploring. In these instances, the dialogue ends once it reaches the predefined conversation length limit. This behavior underscores the importance of restricting the number of interactions with the LLM, as these rare cases can incur significant costs without yielding further speedup improvements.

\subsection{Supplementary Results for RQ1}
\label{sec:rq1_cont}

\begin{figure*}[h]
    \centering
    \includegraphics[width=\textwidth]{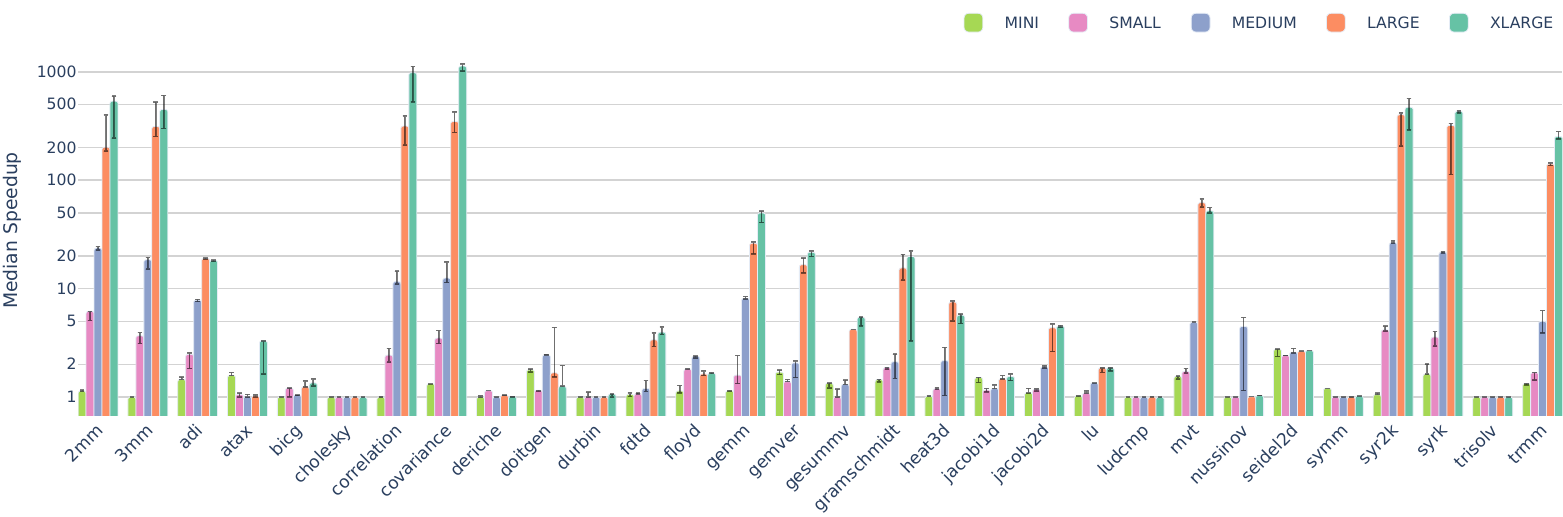}

    \caption{\BestOf{5}{30} speedups per benchmark instance. Each bar represents the median best-of-5 speedup. Error bars indicate the 95\% CIs. }
    \label{fig:collm5_30_speedups} 
    
\end{figure*}

Figure~\ref{fig:collm5_30_speedups} shows the \BestOf{5}{30} speedups on each individual benchmark achieved using the best-of-5 runs after 30 iterations.

Section~3.1 (RQ1) discussed several examples of high-performing optimization schedules discovered by \FrameworkName{}. Due to space constraints, only partial schedules were shown in the main text. Table~\ref{tab:example_schedules} provides the full transformation sequences corresponding to the median speedup runs (\medianAt{30}) for the specific benchmark instances highlighted in RQ1.

\begin{table}[h]
\footnotesize
    \centering
    \caption{Full optimization schedules corresponding to median speedup runs (\medianAt{30}) for benchmark instances discussed in RQ1.}
    \label{tab:example_schedules}
    \setlength{\tabcolsep}{3pt} 
    \begin{tabularx}{\linewidth}{@{} l l >{\raggedright\arraybackslash}X @{}} 
        \toprule
        \textbf{Benchmark Instance} & \textbf{ Speedup} & \textbf{Corresponding Schedule (\medianAt{30})} \\
        \midrule
        \smalltt{correlation\_XLARGE} & 339$\times$ & \smalltt{comp00.Parallelize(L0)} \smalltt{+comp01.Unroll(L-1,4)} \smalltt{+comp02.Parallelize(L0)} \smalltt{+comp03.Unroll(L-1,4)} \smalltt{+comp04.Parallelize(L0)} \smalltt{+comp04.Unroll(L-1,4)} \smalltt{+comp05.Parallelize(L0)}
        \smalltt{+comp07.Tile2D(L1,L2,32,32)} \smalltt{+comp07.Unroll(L2,16)} \\
        \addlinespace 

        \smalltt{trmm\_XLARGE} & 183$\times$ & \smalltt{comp01.Interchange(L0,L1)} \smalltt{+comp01.Tile2D(L1,L2,32,64)} \smalltt{+comp01.Parallelize(L0)} \smalltt{+comp01.Unroll(L1,4)} \\
        \addlinespace

        \smalltt{trmm\_MEDIUM} & 3.6$\times$ & \smalltt{comp00.Tile2D(L0,L1,16,16)} \smalltt{+comp00.Parallelize(L1)} \\
        \addlinespace

        \smalltt{seidel2d\_SMALL} & 2.41$\times$ & \smalltt{comp00.Skew(L1,L2)} \\
        \bottomrule
    \end{tabularx}
\end{table}

\subsection{Supplementary Results for RQ3}
\label{sec:RQ3_cont}
\begin{figure}[h]
    \centering
    \includegraphics[width=\linewidth]{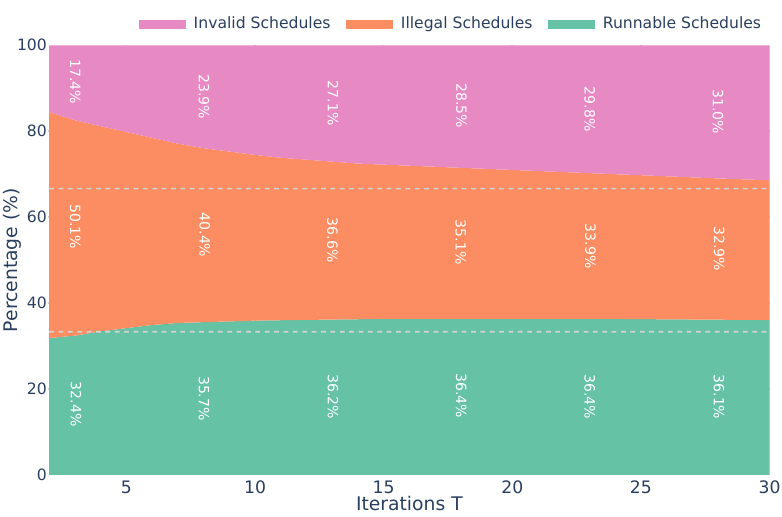}
    \caption{Evolution of schedule viability over dialogue iterations (T=1 to 30). }
    \label{fig:exploration_ratios_over_iterations}
    
\end{figure}

As mentioned in Section~3.1, the ratio of runnable, invalid, and illegal schedules suggested by the LLM changes over the course of the optimization dialogue. Figure~\ref{fig:exploration_ratios_over_iterations} illustrates how these ratios evolve over the first 30 iterations, averaged across all runs and benchmarks using \smalltt{gemini-2.0-flash}.

Initially, illegal schedules are highly prevalent (nearly 60\% at $T=1$), suggesting the LLM often proposes transformations that violate dependencies early on. As the dialogue progresses, the proportion of illegal suggestions tends to decrease, while the proportion of invalid suggestions increases before stabilizing. The runnable ratio gradually improves, converging towards the \textasciitilde 36\% average. This dynamic might indicate the LLM learns to avoid certain dependency-violating patterns based on feedback but may then attempt more complex (and sometimes invalid) combinations.

\subsection{Supplementary Results for RQ4}
\label{sec:RQ4_cont}

Table~\ref{tab:llm_speedups_bestof5} presents the multi-run (\BestOf{5}{T}) geometric mean speedups for the LLMs we tested at various iteration counts ($T$). Best results per column within 5\% tolerance are bolded. 

\begin{table}[h]
\footnotesize

    \centering
    \caption{\BestOf{5}{T} geomean across all benchmarks for different LLMs at various iteration (T).}
    \label{tab:llm_speedups_bestof5}
\setlength{\tabcolsep}{4pt} 
\begin{tabular}{lcccccc}
    \toprule
    LLM & \emph{T}=5 & \emph{T}=10 & \emph{T}=15 & \emph{T}=20 & \emph{T}=25 & \emph{T}=30 \\
        \midrule
        \texttt{gemini-2.0-flash} & 2.39 & 2.86 & \textbf{3.13} & \textbf{3.32} & 3.42 & \textbf{3.54} \\
        \texttt{gemma3 (27B)} & 2.05 & 2.34 & 2.35 & 2.48 & 2.52 & 2.58 \\
        \texttt{gpt-4o} & 2.41 & 2.67 & 2.85 & 2.98 & 3.13 & 3.26 \\
        \texttt{llama3.3 (70B)} & 2.43 & 2.76 & 2.87 & 2.99 & 3.06 & 3.08 \\
        \texttt{gpt-o3-mini} & \textbf{2.68} & \textbf{3.02} & \textbf{3.28} & \textbf{3.44} & \textbf{3.63} & \smalltt{N/E} \\
        \texttt{qwq (32B)} & \textbf{2.61} & 2.83 & 2.93 & 2.94 & 2.94 & 2.94 \\
        \texttt{qwen2.5-coder (32B)} & 2.28 & 2.61 & 2.81 & 2.91 & 2.90 & 3.00 \\
        \texttt{codestral-2501 (22B)} & 1.85 & 1.99 & 2.08 & 2.28 & 2.38 & 2.30 \\
        \bottomrule
    \end{tabular}
\end{table}

Table \ref{tab:llm_efficiency} show the distribution of invalid, illegal, and  runnable schedules for each of the LLMs we tested. 
\begin{table}[h]
\footnotesize
    \centering
    \caption{Percentage breakdown of suggested schedules (averaged up to T=30) for different LLMs.}
    \label{tab:llm_efficiency}\
    \setlength{\tabcolsep}{4pt} 
    \begin{tabular}{lccc}
        \toprule
        LLM & Invalid (\%) & Illegal (\%) & Runnable (\%) \\
        \midrule
        \texttt{gemini-2.0-flash} & 31.4 & 32.5 & 36.1 \\
        \texttt{gemma3 (27B)} & 52.7 & 23.5 & 23.7 \\
        \texttt{gpt-4o} & 31.7 & 30.2 & 38.0 \\
        \texttt{llama3.3 (70B)} & 38.0 & 23.9 & 38.1 \\
        \texttt{gpt-o3-mini} & 31.1 & 28.8 & 40.1 \\
        \texttt{qwq (32B)} & 39.4 & 33.0 & 27.6 \\
        \texttt{qwen2.5-coder (32B)} & 38.1 & 31.4 & 30.5 \\
        \texttt{codestral-2501 (22B)} & 64.4 & 20.3 & 15.3 \\
        \bottomrule
    \end{tabular}
\end{table}

\section{Visualization of Exploration Variability}
\label{app:variability_plots}

As noted in Section~3.1 (RQ1), certain benchmarks exhibit wide confidence intervals (CIs) for their median speedups. This variability stems from the stochastic nature of the LLM's exploration process. Across different runs, \FrameworkName{} can converge towards distinct local optima within the vast transformation search space, leading to different final schedules and associated speedups.

\begin{figure}[h]
    \centering
    \includegraphics[width=\linewidth]{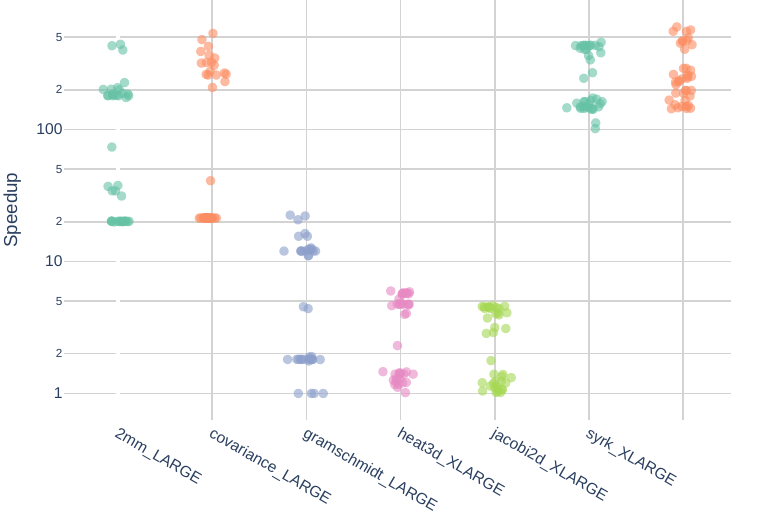} 
    \caption{Distribution of final speedups (\texttt{@30}) for 40 individual runs across selected benchmarks with high variability. Each point represents one run. Clustering indicates convergence to different local optima.}
    \label{fig:app_strip_plot_variability}
\end{figure}

Figure~\ref{fig:app_strip_plot_variability} provides direct evidence of this phenomenon. It displays the distribution of speedups achieved by each individual run (40 runs in total per benchmark) after 30 iterations (\texttt{@30}) for seven benchmarks that showed notably high CIs. The distinct clustering of points for each benchmark visually confirms the presence of multi-modal speedup distributions. Each cluster likely represents a different set of effective optimization schedules discovered by the LLM across various runs.

\begin{figure}[h]
    \centering
    \includegraphics[width=\linewidth]{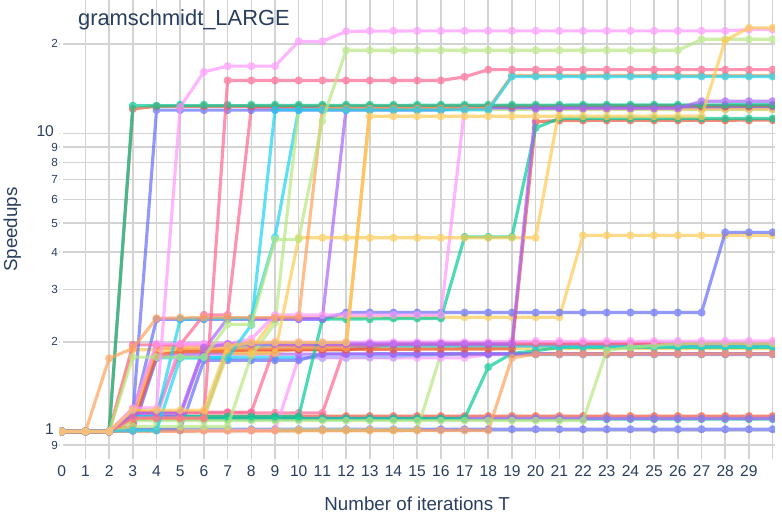} 
    \caption{Evolution of speedup over iterations (T=0 to T=30) for 40 individual runs on \smalltt{gramschmidt\_LARGE}. Distinct trajectories show different exploration paths.}
    \label{fig:app_line_plot_gramschmidt}
\end{figure}

\begin{figure}[h]
    \centering
    \includegraphics[width=\linewidth]{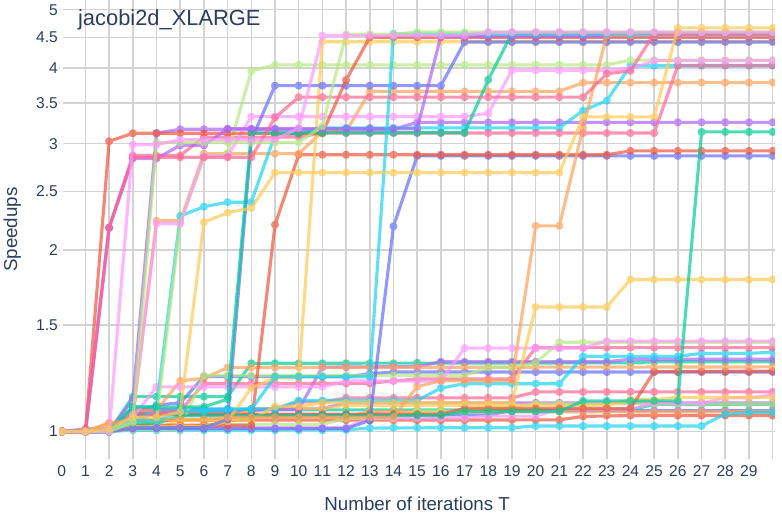} 
    \caption{Evolution of speedup over iterations (T=0 to T=30) for 40 individual runs on \smalltt{jacobi2d\_XLARGE}, further illustrating varied exploration paths.}
    \label{fig:app_line_plot_jacobi2d}
\end{figure}

Further insight into how these different optima are reached during the exploration is provided in Figures \ref{fig:app_line_plot_gramschmidt} and \ref{fig:app_line_plot_jacobi2d}. These plots track the evolution of the best speedup found so far over the 30 iterations for each individual run, specifically for the \smalltt{gramschmidt\_LARGE} and \smalltt{jacobi2d\_XLARGE} instances, respectively. The plots clearly show distinct "bundles" or trajectories of lines. Different runs not only reach different final speedups but often follow divergent paths throughout the optimization dialogue, reinforcing the idea that the LLM explores and settles into different regions of the optimization space.

Together, these visualizations illustrate that the observed variance in speedups, particularly for certain benchmarks, is a direct consequence of the LLM exploring different pathways and converging to multiple, distinct performance optima. This underscores the value of the multi-run strategy (\BestOf{K}{T}) discussed in the main results for increasing the probability of finding one of the better-performing optima.


\end{document}